\documentclass{aastex63}
\usepackage{threeparttable}
\usepackage{amsmath,bm}
\usepackage{longtable}
\usepackage{tabu}
\usepackage{multirow}

\submitjournal{ApJS}
\shorttitle{LAMOST high velocity star}
\shortauthors{Li Yin-Bi et al.}

\begin{document}

\title{591 high velocity stars in the Galactic halo selected from LAMOST DR7 and Gaia DR2}

\correspondingauthor{A-Li Luo, You-Jun Lu, Gang Zhao}
\email{lal@bao.ac.cn, luyj@bao.ac.cn, gzhao@bao.ac.cn}

\author[0000-0001-7607-2666]{Yin-Bi Li}
\affiliation{CAS Key Laboratory of Optical Astronomy, National Astronomical Observatories, Chinese Academy of Sciences, Beijing 100101, China}

\author[0000-0001-7865-2648]{A-Li Luo$^{*}$}
\affiliation{CAS Key Laboratory of Optical Astronomy, National Astronomical Observatories, Chinese Academy of Sciences, Beijing 100101, China}
\affiliation{School of Astronomy and Space Science, University of Chinese Academy of Sciences, Beijing 100049, China}


\author[0000-0002-1310-4664]{You-Jun Lu$^{*}$}
\affiliation{CAS Key Laboratory of Optical Astronomy, National Astronomical Observatories, Chinese Academy of Sciences, Beijing 100101, China}
\affiliation{School of Astronomy and Space Science, University of Chinese Academy of Sciences, Beijing 100049, China}

\author{Xue-Sen Zhang}
\affiliation{ExtantFuture (Beijing) Technology Co., Ltd, Beijing 100102, People$'$ Republic of China}

\author[0000-0002-2577-1990]{Jiao Li}
\affiliation{CAS Key Laboratory of Optical Astronomy, National Astronomical Observatories, Chinese Academy of Sciences, Beijing 100101, China}
\affiliation{Yunnan observatories, Chinese Academy of Sciences, Kunming 650011, China}

\author[0000-0001-6767-2395]{Rui Wang}
\affiliation{CAS Key Laboratory of Optical Astronomy, National Astronomical Observatories, Chinese Academy of Sciences, Beijing 100101, China}

\author{Fang Zuo}
\affiliation{CAS Key Laboratory of Optical Astronomy, National Astronomical Observatories, Chinese Academy of Sciences, Beijing 100101, China}
\affiliation{School of Astronomy and Space Science, University of Chinese Academy of Sciences, Beijing 100049, China}

\author[0000-0002-5818-8769]{Maosheng Xiang}
\affiliation{Max-Planck Institute for Astronomy, Konigstuhl 17, D-69117 Heidelberg, Germany}

\author[0000-0001-5082-9536]{Yuan-Sen Ting}
\affiliation{Institute for Advanced Study, Princeton, NJ 08540, USA}
\affiliation{Department of Astrophysical Sciences, Princeton University, Princeton, NJ 08540, USA}
\affiliation{Observatories of the Carnegie Institution of Washington, 813 Santa Barbara Street, Pasadena, CA 91101, USA}
\affiliation{Research School of Astronomy \& Astrophysics, Australian National University, Cotter Rd., Weston, ACT 2611, Australia}

\author{T.Marchetti}
\affiliation{Leiden Observatory, Leiden University, PO Box 9513 2300 RA Leiden, the Netherlands}

\author{Shuo Li}
\affiliation{CAS Key Laboratory of Optical Astronomy, National Astronomical Observatories, Chinese Academy of Sciences, Beijing 100101, China}
\affiliation{School of Astronomy and Space Science, University of Chinese Academy of Sciences, Beijing 100049, China}

\author{You-Fen Wang}
\affiliation{CAS Key Laboratory of Optical Astronomy, National Astronomical Observatories, Chinese Academy of Sciences, Beijing 100101, China}

\author[0000-0003-1454-1636]{Shuo Zhang}
\affiliation{Department of Astronomy, School of Physics, Peking University, Beijing 100871, China}
\affiliation{Kavli Institute for Astronomy and Astrophysics, Peking University, Beijing 100871, China}

\author{Kohei Hattori}
\affiliation{Department of Astronomy, University of Michigan, 1085 S. University Ave., Ann Arbor, MI 48109, USA}

\author{Yong-Heng Zhao}
\affiliation{CAS Key Laboratory of Optical Astronomy, National Astronomical Observatories, Chinese Academy of Sciences, Beijing 100101, China}
\affiliation{School of Astronomy and Space Science, University of Chinese Academy of Sciences, Beijing 100049, China}

\author[0000-0002-7727-1699]{Hua-Wei Zhang}
\affiliation{Department of Astronomy, School of Physics, Peking University, Beijing 100871, China}
\affiliation{Kavli Institute for Astronomy and Astrophysics, Peking University, Beijing 100871, China}

\author[0000-0002-8980-945X]{Gang Zhao$^{*}$}
\affiliation{CAS Key Laboratory of Optical Astronomy, National Astronomical Observatories, Chinese Academy of Sciences, Beijing 100101, China}
\affiliation{School of Astronomy and Space Science, University of Chinese Academy of Sciences, Beijing 100049, China}

\begin{abstract} 

In this paper, we report 591 high velocity star candidates (HiVelSCs) selected from over 10 million spectra of the data release seven (DR7) of the Large Sky Area Multi-object Fiber Spectroscopic Telescope and the second Gaia data release, with three-dimensional velocities in the Galactic rest-frame larger than $445$ km~s$^{-1}$. We show that at least 43 HiVelSCs are unbound to the Galaxy with escape probabilities larger than 50$\%$, and this number decreases to eight if the possible parallax zero-point error is corrected. Most of these HiVelSCs are metal-poor and slightly $\alpha$-enhanced inner halo stars. Only 14$\%$ of them have [Fe/H]~$>$~-1, which may be the metal-rich ``in situ'' stars in halo formed in the initial collapse of the Milky Way or metal-rich stars formed in the disk or bulge but kinematically heated. The low ratio of 14$\%$ implies that the bulk of stellar halo was formed from the accretion and tidal disruption of satellite galaxies. In addition, HiVelSCs on the retrograde orbits have a slightly lower metallicities on average compared with these on the prograde orbits, meanwhile metal-poor HiVelSCs with [Fe/H]~$<$~-1 have an even faster mean retrograde velocity compared with metal-rich HiVelSCs. To investigate the origins of HiVelSCs, we perform orbit integrations and divide them into four types, i.e., hypervelocity stars, hyper-runaway stars, runaway stars and fast halo stars. A catalog for these 591 HiVelSCs, including radial velocities, atmospheric parameters, Gaia astrometric parameters, spatial positions, and velocities, etc., is  available at \url{http://paperdata.china-vo.org/LYB/lamostdr7_gaiadr2_hvs_591.csv}.



\end{abstract}


\section{Introduction} \label{sec:intro}
High velocity stars are moving fast, and they mark the presence of
extreme dynamical and astrophysical processes, especially when a star approaches
or even exceeds the escape velocity of the Galaxy at its position \citep{1988Natur.331..687H, 2003ApJ...599.1129Y, 2006ApJ...653.1194B, 2009ApJ...691L..63A, 2008MNRAS.383...86O, 2015MNRAS.454.2677C, 2019MNRAS.490..157M}.
These stars provide probes for a wide range of Galactic science, on scales from a few
parsecs near the central Massive Black Hole (MBH) to the distant Galactic halo.
They can provide insights to the dynamical  mechanisms that produce their extreme velocities \citep{2010ApJ...722.1744Z, 2015MNRAS.447.2046H},
and for example, their  distributions in space and velocity can
reveal the existence of a binary MBH \citep{2010ApJ...722.1744Z, 2015ARA&A..53...15B, 2018ApJ...856...92F, 2018MNRAS.475.4986F}. They are powerful tracers
used to probe the mass distribution of the Galaxy since they travel large distances across
the Galaxy \citep{2005ApJ...634..344G, 2008ApJ...680..312K}, and their trajectories can also be
used to probe the shape of the dark matter halo of the Galaxy
\citep{2008ApJ...680..312K, 2007MNRAS.379.1293Y}.

In general, there are four subclasses for high velocity stars including ``hypervelocity star" (HVS),
``runaway star" (RS), ``hyper-runaway star" (HRS), and ``fast halo star'' (OUT), and they have different
origins. The fastest stars in our Galaxy are HVSs, and their proposed ejection
mechanisms include: the tidal breakup of binary stars by a single MBH in
the Galactic center (GC) \citep{1988Natur.331..687H, 2003ApJ...599.1129Y, 2006ApJ...653.1194B}; single star encounters with
a binary MBH \citep{2003ApJ...599.1129Y, 2006ApJ...651..392S, 2007MNRAS.379L..45S}; single star encounters
with a cluster of stellar mass black holes around the MBH \citep{2008MNRAS.383...86O}; the interaction
between a globular cluster with a single or binary MBH in the GC \citep{2015MNRAS.454.2677C, 2016MNRAS.458.2596F}.
RSs are the second subclass of high velocity stars, and they are thought to have formed in the disk and then were
ejected into the halo. RSs can be produced through two main formation mechanisms: supernova explosions
in stellar binary systems  \citep[e.g.,][]{1961BAN....15..265B, 2009MNRAS.396..570G, 2009A&A...508L..27W, 2013A&A...559A..94W};
dynamical  interactions due to multi-body encounters in dense stellar systems \citep[e.g.,][]{2009MNRAS.396..570G, 2009ApJ...706..925B}.
If a RS has such an extremely high velocity that it can escape from the Galaxy, it belongs to the third subclass of high velocity stars, i.e., to the HRS \citep{2012ApJ...751..133P, 2015ARA&A..53...15B, 2018AJ....156...87L}. The last subclass of high velocity stars is the fast halo stars, and they can be produced by the tidal interactions of dwarf galaxies near the GC \citep{2009ApJ...691L..63A}, or from other galaxies in the Local group \citep{2008MNRAS.386.1179S, 2009ApJ...707L..22T}, for example from the center of the Large Magellanic Cloud \citep[LMC, ][]{2016ApJ...825L...6B, 2017MNRAS.469.2151B}.

The first hypervelocity star (HVS1) was serendipitously discovered by \citet{2005ApJ...622L..33B}, and
it is a late-B type star with a heliocentric distance $\sim$71 kpc and radial velocity ($RV$) $\sim$853
km s$^{-1}$. Since then, the number of high velocity star candidates (HiVelSCs) has ballooned, and there are
nearly 500 candidates in the literature before Gaia DR2 \citep{2005A&A...444L..61H, 2005ApJ...634L.181E, 2006ApJ...647..303B, 2009ApJ...690.1639B, 2012ApJ...751...55B, 2014ApJ...787...89B, 2008A&A...483L..21H, 2009ApJ...697.1543K,
2009JPhCS.172a2009T, 2012ApJ...744L..24L, 2014ApJ...785L..23Z, 2014ApJ...789L...2Z, 2015RAA....15.1364L, 2015ApJ...813...26F, 2017ApJ...847L...9H, 2017Sci...357..680V}. Most of these stars are late type high proper motion stars, and about two dozens of them are faint and blue stars in the halo and
were classified as HVSs based only on their extreme $RV$s.

After Gaia DR2 release, new high velocity stars were searched for with more precise astrometric parameters
\citep{2018AJ....156...87L, 2018ApJ...865...15S, 2018ApJ...866..121H, 2018A&A...615L...5I, 2018ApJ...868...25B,
2019A&A...628L...5I, 2019MNRAS.489.1489R, 2019A&A...627A.104D, 2020A&A...638A.122C, 2020MNRAS.491.2465K, 2020RAA....20...42L}.
For example, \citet{2018AJ....156...87L} found a new late-B type hyper-runaway star from the Large Sky Area Multi-object Fiber Spectroscopic
Telescope (LAMOST) with Gaia DR2 proper motions, and it has a total Galactocentric velocity of $\sim$ 586 km~s$^{-1}$. \citet{2018ApJ...866..121H} reported the discovery of 30 old metal-poor stars with extreme velocities ($>$ 480 km s$^{-1}$) in Gaia DR2, up to three of these stars were purported to be ejected from the LMC, and one or two stars originated from the GC. \citet{2019MNRAS.490..157M} found 20 stars that have
probabilities $>$ 80$\%$ of being unbound from the Galaxy, seven of them are hyper-runaway stars, and other 13 stars may not have their origins in the
Galaxy. \citet{2018ApJ...865...15S} found three hyper-runaway white dwarfs in Gaia DR2, which have total
Galactocentric velocities between 1000 and 3000 km s$^{-1}$. Except for these newly discovered high velocity stars, previously
known HVSs were also revisited to further investigate their possible origins by tracing their orbits back in
time with new Gaia astrometric parameters \citep{2018ApJ...866...39B, 2018MNRAS.479.2789B, 2019MNRAS.483.2007E, 2018A&A...615L...5I, 2020A&A...637A..53K}.
For example, \citet{2019MNRAS.483.2007E} reanalyzed 26 previously known HVSs and found
that the third hypervelocity star (HE 0437-5439 or HVS3) is likely to be coming almost from the center of the LMC, and \citet{2018MNRAS.479.2789B} found that almost all previously known late type high velocity stars
are likely bound to the Milky Way and only one late-type object (LAMOST J115209.12+120258.0) is unbound from the
Galaxy \citep{2018AJ....156...87L}.

\citet{2018ApJ...863...87D} and \citet{2018ApJ...869L..31D} successively searched
for HiVelSCs from the early data version of LAMOST and Gaia, and the second work only focuses on spectra also having Gaia $RV$s (LAMOST provides $RV$s data for all stellar spectra, but Gaia provides these data only for a small fraction of spectral types), which is a sample smaller than that corresponding to the LAMOST data they used. As mentioned in \citet{2019MNRAS.486.2618B} and \citet{2020RAA....20...42L}, $RV$s of Gaia DR2 could be spurious if there exists light contamination from a nearby bright star. To avoid this problem, we only use LAMOST $RV$s to estimate spatial velocities, and search for HiVelSCs from the latest data version of LAMOST (data release seven) and Gaia (data release two). Besides, different target selection method, and distance and velocity estimation methods are adopted in this work, which leads to a different search result of HiVelSCs.

Up to now, only a few studies have used both the chemical and kinematic information of high velocity stars to determine where they were produced and how did they achieve such high velocities \citep{2015MNRAS.447.2046H, 2015Sci...347.1126G, 2018ApJ...863...87D, 2018ApJ...869L..31D, 2018A&A...620A..48I, 2019MNRAS.490..157M}. In this paper, we use the chemistry and kinematics simultaneously to investigate the possible origins and stellar populations for high velocity stars, and find a few possible ``in situ'' halo stars. We also analyze the kinematic properties for metal-poor and metal-rich stars, respectively, and the chemical properties for prograde and retrograde stars.

The paper is organized as follows. In Section 2, we describe the method that we used to search for HiVelSCs. In Section 3, we analyze the
spatial position and velocity distributions of our HiVelSCs, their stellar population, and their chemical and kinematic properties. In Section 4, we
calculate orbit parameters for HiVelSCs, and investigate their possible origins. In Section 5, we discuss the impact of a -0.029 mas global parallax zero-point, and the results that adopt more conservative criteria to select HiVelSCs. Finally, the conclusions are presented in Section 6.

\section{Data} \label{sec:data}

\subsection{LAMOST and Gaia} \label{sec:lamost&gaia}

LAMOST is a 4m quasi-meridian reflective Schmidt telescope, which is equipped with 4000 fibers and can observe up to 4000
targets per exposure simultaneously
\citep{1996ApOpt..35.5155W, 2004ChJAA...4....1S, 2006ChJAA...6..265Z, 2012RAA....12..723Z, 2012RAA....12.1243L, 2012RAA....12.1197C}.
From 2011 to 2018, LAMOST has completed its first stage low resolution spectroscopic survey ($R$~$\sim$~1800), and obtained
more than 9 million spectra \footnote{http://dr5.lamost.org/}. Since October 2018, LAMOST started the second stage survey (LAMOST II),
which contains both low- and medium-resolution spectroscopic surveys. LAMOST II takes about 50$\%$ nights to continue the previous
low-resolution survey, and the other 50$\%$ nights (bright/gray nights) for medium-resolution survey \citep{2020arXiv200507210L}.
In March 2020, DR7 of Low-Resolution Spectroscopic Survey provided 10,608,416 low resolution spectra \footnote{http://dr7.lamost.org/},
and these spectra cover the wavelength range of 3690$-$9100 \AA~with a resolution of $R$~$\sim$~1800 respectively at 4750
\AA~(blue) and 7350 \AA~(red). The LAMOST Stellar Parameter Pipeline (LASP) estimates atmospheric parameters and $RV$,
and it has an accuracy of about 150 K, 0.25 dex, 0.15 dex, 5.0 km s$^{-1}$ for the effective temperature ($T_{\rm eff}$),
surface gravity (log~$g$), metallicity ([Fe/H]) and $RV$, respectively \citep{2015RAA....15.1095L, 2015MNRAS.448..822X},
for spectra of signal to noise ratios higher than 10.

Gaia is a space-based mission in the science programme of the European Space Agency (ESA) launched in 2013,
and its main aim is to measure astrometric parameters of stars, and to understand the formation, structure,
and evolution of the Milky Way \citep[see review from][]{2016ARA&A..54..529B}. On 2018 April 25, Gaia delivered its second date
release \citep[Gaia DR2, ][]{2016A&A...595A...1G, 2018A&A...616A...1G}, and it provides precise positions
($\alpha$, $\delta$), proper motions ($\mu_{\alpha^{*}}$, $\mu_{\delta}$), parallaxes ($\varpi$) and photometries
for over 1.3 billion stars brighter than magnitude 21 \citep{2018A&A...616A...1G, 2018A&A...616A..14G}.
The median uncertainty in parallax and position (at the reference epoch J2015.5) is about 0.04 mas at $G <$ 14 mag,
0.1 mas at $G$ = 17 mag, and 0.7 mas at $G$ = 20 mag, and the corresponding uncertainties of the proper motion
components are 0.05, 0.2, and 1.2 mas yr$^{-1}$, respectively \citep{2018A&A...616A...2L}.

\subsection{Distance and Total Velocity Determination} \label{sec:distance_velocity}

Using the five astrometric parameters in Gaia DR2 catalogue and $RV$ in LAMOST catalogue,
Galactocentric distance ($r_{\rm GC}$) and total velocity ($V_{\rm GC}$)
are computed based on the following assumptions: 1) the distance between the Sun and the GC is 
$d_{\odot}$ = 8.2 kpc, and the Sun has an offset above the stellar disk of $z_{\odot}$ = 25 pc
\citep{2016ARA&A..54..529B}; 2) the motion of the local standard of rest (LSR) is 238 km s$^{-1}$,
and the velocity of the Sun with respect to the LSR is [U$_{\odot}$, V$_{\odot}$,
W$_{\odot}$] = [14.0, 12.24, 7.25] km s$^{-1}$ \citep{2010MNRAS.403.1829S, 2012MNRAS.427..274S, 2016ARA&A..54..529B};
3) the distribution of equator coordinates, parallax and proper motions is a multivariate
Gaussian with a mean vector m = [$\alpha$, $\delta$, $\varpi$, $\mu_{\alpha^{*}}$, $\mu_{\delta}$], and a

covariance matrix:
\begin{align}\label{equ:covariance-matrix}
    \Sigma =
\begin{pmatrix}
\sigma^{2}_{\rm \alpha} & \quad\sigma_{\rm \alpha}\sigma_{\rm \delta}\rho(\alpha, \delta) & \quad\sigma_{\rm \alpha}\sigma_{\rm \varpi}\rho(\alpha, \varpi) & \quad\sigma_{\rm \alpha}\sigma_{\rm \mu_{\alpha^{*}}}\rho(\alpha, \mu_{\alpha^{*}}) & \quad\sigma_{\rm \alpha}\sigma_{\rm \mu_{\delta}}\rho(\alpha, \mu_{\delta}) \\
\sigma_{\rm \delta}\sigma_{\rm \alpha}\rho(\delta, \alpha) & \quad\sigma^{2}_{\rm \delta} & \quad\sigma_{\rm \delta}\sigma_{\rm \varpi}\rho(\delta, \varpi) & \quad\sigma_{\rm \delta}\sigma_{\rm \mu_{\alpha^{*}}}\rho(\delta, \mu_{\alpha^{*}}) & \quad\sigma_{\rm \delta}\sigma_{\rm \mu_{\delta}}\rho(\delta, \mu_{\delta}) \\
\sigma_{\rm \varpi}\sigma_{\rm \alpha}\rho(\varpi, \alpha) & \quad\sigma_{\rm \varpi}\sigma_{\rm \delta}\rho(\varpi, \delta) & \quad\sigma^{2}_{\rm \varpi} & \quad\sigma_{\rm \varpi}\sigma_{\rm \mu_{\alpha^{*}}}\rho(\varpi, \mu_{\alpha^{*}}) & \quad\sigma_{\rm \varpi}\sigma_{\rm \mu_{\delta}}\rho(\varpi, \mu_{\delta}) \\
\sigma_{\rm \mu_{\alpha^{*}}}\sigma_{\rm \alpha}\rho(\mu_{\alpha^{*}}, \alpha) & \quad\sigma_{\rm \mu_{\alpha^{*}}}\sigma_{\rm \delta}\rho(\mu_{\alpha^{*}}, \delta) & \quad\sigma_{\rm \mu_{\alpha^{*}}}\sigma_{\rm \varpi}\rho(\mu_{\alpha^{*}}, \varpi) & \quad\sigma^{2}_{\rm \mu_{\alpha^{*}}} & \quad\sigma_{\rm \mu_{\alpha^{*}}}\sigma_{\rm \mu_{\delta}}\rho(\mu_{\alpha^{*}}, \mu_{\delta}) \\
\sigma_{\rm \mu_{\delta}}\sigma_{\rm \alpha}\rho(\mu_{\delta}, \alpha) & \quad\sigma_{\rm \mu_{\delta}}\sigma_{\rm \delta}\rho(\mu_{\delta}, \delta) & \quad\sigma_{\rm \mu_{\delta}}\sigma_{\rm \varpi}\rho(\mu_{\delta}, \varpi) & \quad\sigma_{\rm \mu_{\delta}}\sigma_{\rm \mu_{\alpha^{*}}}\rho(\mu_{\delta}, \mu_{\alpha^{*}}) & \quad\sigma^{2}_{\rm \mu_{\delta}}
\end{pmatrix}
\end{align}

where $\rho$($i$, $j$) is the correlation coefficient between the astrometric parameters $i$ and $j$,
and is provided by the Gaia DR2. LAMOST $RV$ is uncorrelated to the astrometric parameters,
thus we assume it follows a Gaussian distribution, which is centered on $RV$ and with a standard deviation of
$\sigma_{\rm RV}$ \citep[uncertainty on $RV$, ][]{2019MNRAS.490..157M}.

To estimate $r_{\rm GC}$ and $V_{\rm GC}$, the Monte Carlo (MC) method is applied.
For each star, 1000 MC samplings on its five astrometric parameters and $RV$ are performed,
and 1000 $r_{\rm GC}$ and $V_{\rm GC}$ can be obtained according to above assumptions. The Galactic Cartesian
coordinate system adopted here is centered on the GC. The X-axis points from the
Sun to the GC, the Y-axis points in the direction of Galactic rotation, and
the Z-axis points toward the Northern Galactic Pole \citep{1987AJ.....93..864J, 2012ApJ...744L..24L}. We use
the median (50th percentile) and the 16th and 84th percentiles as the computed
parameters and their lower and upper uncertainties, respectively \citep{2019MNRAS.490..157M, 2019A&A...627A.104D}.

\citet{2018A&A...616A...2L} pointed that Gaia DR2 parallax has a global zero-point of -0.029 mas, which indicates that the parallaxes
are underestimated and hence the distances are overestimated if the usual inverse relationship is adopted. However, both
\citet{2018A&A...616A...2L} and \citet{2018A&A...616A..17A} explicitly discouraged a global zero-point correction, particularly in
cases where the sample is not well distributed over the entire sky. In addition, \citet{2018A&A...616A..17A} showed that the parallax
offset is partly dependent on the scanning pattern of Gaia, which makes it a function of the coordinates. Taking these into account,
we do not correct the parallaxes zero-point here, and discuss the changes on distances and velocities
if the zero-point correction were to be considered in Section~\ref{sec:discussion&conclusion}.

\subsection{Sample Selection} \label{sec:sample}

Using the parallax in Gaia DR2 catalogue, the heliocentric distance can be determined just by inverting the parallax:
$d = 1 / \varpi$. However, \citet{2015PASP..127..994B} discussed that this naive approach fails for
non-positive parallaxes, and can induce strong biases for fractional parallax errors, i.e.,
$f_{\rm \varpi} = \sigma_{\varpi} / \varpi$, larger than about 20$\%$. In this paper, we only focus on
stars with positive parallaxes ($\varpi > 0$) and smaller fractional parallax errors of $f_{\rm \varpi} \leq 0.2$
(the ``low-f'' samples).


Using the TOPCAT \footnote{http://www.star.bris.ac.uk/~mbt/topcat/}, cross-match of LAMOST DR7 with Gaia DR2 were
performed with a radius of 5 arc sec, and outputs a catalogue (LAMOST-Gaia hereafter) consisting of over 10 million entries,
which includes both LAMOST and Gaia parameters, for example, $RV$, five Gaia astrometric parameters, magnitudes,
and correlation coefficients. In LAMOST-Gaia catalogue, there are over 8.48 million ``low-f'' entries, and their
$r_{\rm GC}$ and $V_{\rm GC}$ are estimated as described in Section~\ref{sec:distance_velocity}. Seven
Galactic potential models \citep[][hereinafter Paczynski+1990, Gnedin+2005, Xue+2008, Koposov+2010, Kenyon+2014, MWPotential2014, and Watkins+2019]{1990ApJ...348..485P, 2005ApJ...634..344G, 2008ApJ...684.1143X, 2010ApJ...712..260K, 2014ApJ...793..122K, 2015ApJS..216...29B, 2019ApJ...873..118W} are adopted here to estimate the escape velocities, and 31,440 samples are picked out with the condition of either escaping from our Galaxy
under at least one potential model or bound but at least with total velocities of $V_{\rm GC} \geq 450$ km s$^{-1}$ as used in \citet{2019MNRAS.490..157M} to
select HivelSCs. We then check the spectral qualities and only retain spectra with the highest signal to noise ratios for stars with multiple observations, and 1761 stars are left. In previous steps, we use the $RV$ provided by the LAMOST 1D pipeline ($RV_{\rm 1D}$) to calculate the total velocity, and did not consider the reliability of RV. To select candidates with reliable $RV$s, we inspect the spectra for 1761 stars to check whether the spectral line shift is consistent with
$RV_{\rm 1D}$, and 591 HiVelSCs with reliable $RV_{\rm 1D}$s are finally selected.

In the public stellar parameter catalog of LAMOST, $RV$ and atmospheric parameters, radial velocity, and their errors were determined by the LASP using the method mentioned in \citet{Luodr6}. This work not only consider the calculation error of the LASP but also include the errors introduced by observation and data processing. Thus, the calculation of LASP $RV$ ($RV_{\rm LASP}$) error takes into account more factors, which may make the outcome closer to the true $RV$ errors. To be consistent with the public LAMOST parameter catalog and use more reasonable RV uncertainties, we adopt the $RV_{\rm LASP}$ as our final $RV$s for 591 HiVelSCs.

\begin{figure*}
\begin{center}
\includegraphics[scale=0.4,angle=0]{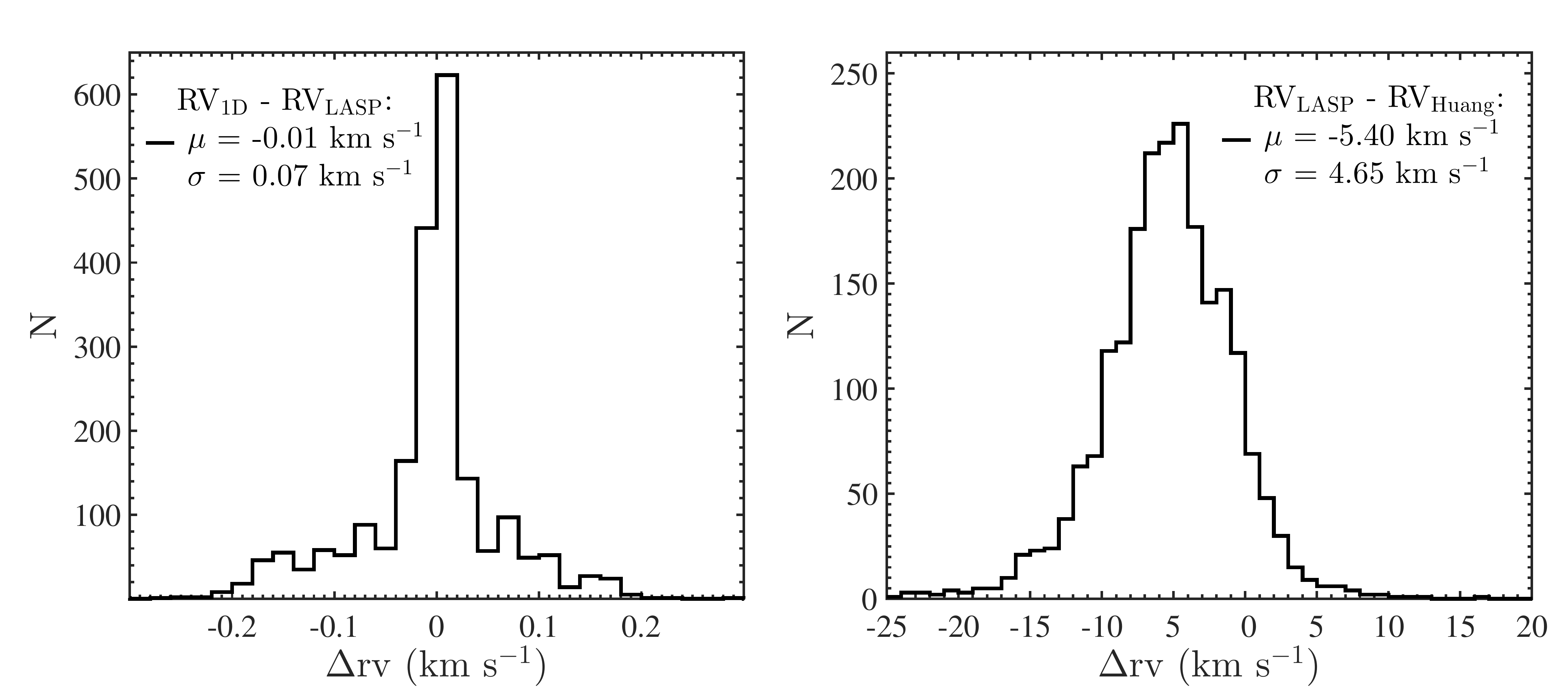}
\caption{Cross-match between a catalog of radial velocity standard stars provided by \citet{2018AJ....156...90H} and LAMOST DR7
catalog were performed, and obtained 2125 common stars. These stars were used to investigate the consistency
ofradial velocities given by the LAMOST 1D pipeline ($RV_{\rm 1D}$) and by the LAMOST Stellar Parameter Pipeline ($RV_{\rm LASP}$) and the radial velocity zero-point, and about 93$\%$ of their spectra  have signal to noise ratios larger than 40. Left panel: the difference between $RV_{\rm 1D}$ and $RV_{\rm LASP}$. Right panel: the difference between $RV_{\rm 1D}$ and radial velocities in the radial velocity standard star catalog ($RV_{\rm Huang}$). A mean difference of $\mu = -0.01$ km s$^{-1}$ represents the $RV_{\rm 1D}$ is consistent with $RV_{\rm LASP}$, which is used to recalculate Galactocentric total velocity ($V_{\rm GC}$) for 592 high velocity star candidates (HiVelSCs), and $RV_{\rm LASP}$ is shown in our high velocity star candidate catalog instead of $RV_{\rm 1D}$. A mean difference $\mu = -5.40$ km s$^{-1}$ implies the zero-point of $RV_{\rm LASP}$ is $-$5.40 km s$^{-1}$, which is used to perform the zero-point correction for 592 HiVelSCs.\label{fig:rv_calibrate}}
\end{center}
\end{figure*}

Then we correct the zero-point of $RV_{\rm LASP}$ for all HiVelSCs using a catalog
of 18080 radial velocity standard stars \citep{2018AJ....156...90H} selected from the APO Galactic Evolution Experiment data, and it is
also used to investigate the consistency of $RV_{\rm 1D}$ and $RV_{\rm LASP}$. We cross-match the catalog of radial velocity standard stars with LAMOST DR7,
and obtain 2125 common stars. For these stars, the distribution of difference between $RV_{\rm 1D}$ and $RV_{\rm LASP}$ is shown in
the left panel of Figure~\ref{fig:rv_calibrate}, and we can see that these stars yield a mean difference of $\mu = -0.01$ km s$^{-1}$ and a standard
deviation of $\sigma = 0.07$ km s$^{-1}$, which represents that $RV_{\rm 1D}$ is well consistent with $RV_{\rm LASP}$. The distribution of difference between $RV_{\rm LASP}$ and $RV$ in the catalog of radial velocity standard stars ($RV_{\rm Huang}$) for these common stars is shown in the right panel of Figure~\ref{fig:rv_calibrate}, and we see that there exists a mean difference of $\mu = -5.40$ km s$^{-1}$ and a standard deviation of $\sigma = 4.65$ km s$^{-1}$,
which means that the zero-point of $RV_{\rm LASP}$ is $-$5.4 km s$^{-1}$. After a zero-point correction of $-$5.4 km s$^{-1}$ for $RV_{\rm LASP}$, we recalculate the spatial velocities for 591 HiVelSCs, and find that the recalculated median $V_{\rm GC}$ are all larger than 445 km s$^{-1}$.

\subsection{High Velocity Star Candidates} \label{sec:candidates}

After the above steps in Section~\ref{sec:sample}, 591 HiVelSCs are finally selected from the LAMOST-Gaia catalogue, and only 14 of them were
reported in the current version of the open fast stars catalog (OFSC) \footnote{https://faststars.space/}
\citep{2018MNRAS.479.2789B}, which contains a collection of 558 HiVelSCs in the literature. \citet{2018ApJ...863...87D} and \citet{2018ApJ...869L..31D} reported 16 (Du1) and 24 (Du2) high velocity stars, respectively, which were both found from LAMOST and Gaia with different data, and nine of them in total were also reported in the OFSC. We cross-match 591 HiVelSCs with Du1 and Du2 in a radius of 3 arc sec, respectively, and find that
12 out of our 591 HiVelSCs were already reported in Du2.

We construct a catalogue including 93 columns to list various parameters for these 591 HiVelSCs, i.e., ``LAMOSTDR7-GAIADR2-HiVelSC'', and Table~\ref{tab:general_catalog} explains each column of this table in detail. The 4th to 21st columns of Table~\ref{tab:general_catalog} are listed in Table~\ref{tab:basic} for the 20 fastest HiVelSCs, such as equatorial coordinate, Gaia parallax and magnitude, and the 26th to 61st columns are listed in Table~\ref{tab:kinematic} also for the 20 fastest HiVelSCs.

\begin{longrotatetable}
\begin{deluxetable*}{ccccccccccccc}
\tablecaption{Basic parameters for the 20 fastest High velocity star candidates (HiVelSCs)\label{tab:basic}}
\tablewidth{0pt}
\tablehead{
\colhead{ID} & \colhead{R.A.} & \colhead{decl.} & \colhead{$S/N$\_r$^{a}$} & \colhead{Class$^{b}$} & \colhead{$RV_{\rm LASP}^{c}$} & \colhead{pmra$^{d}$} &
\colhead{pmdec$^{d}$} & \colhead{parallax$^{e}$} & \colhead{$G^{f}$} & \colhead{$G_{\rm BP}^{f}$} & \colhead{$G_{\rm RP}^{f}$} & \colhead{astrometric\_flag$^{g}$} \\
\colhead{} & \colhead{(deg)} & \colhead{(deg)} & \colhead{} & \colhead{} & \colhead{(km s$^{-1}$)} & \colhead{(mas yr$^{-1}$)} & \colhead{(mas yr$^{-1}$)} &
\colhead{(mas)} & \colhead{(mag)} & \colhead{(mag)} & \colhead{(mag)} & \colhead{}
}
\startdata
Hivel1 & 240.3374980 & 41.16681800 & 158 & G7 & -179$\pm$5 & -25.759$\pm$0.025 & -9.745$\pm$0.040 & 0.118$\pm$0.016 & 13.01 & 13.00 & 15.21 & 0  \\
Hivel2 & 193.4372560 & 55.05813200 & 29 & F9 & -225$\pm$14 & -23.742$\pm$0.042 & -39.545$\pm$0.041 & 0.197$\pm$0.029 & 15.63 & 13.19 & 13.89 & 1  \\
Hivel3 & 102.4840100 & 46.83601200 & 33 & K5 & -75$\pm$6 & 142.070$\pm$0.091 & -285.151$\pm$0.084 & 1.345$\pm$0.092 & 14.39 & 16.20 & 16.92 & 0  \\
Hivel4 & 171.7747380 & 42.40792700 & 135 & A1V & 109$\pm$4 & -39.672$\pm$0.054 & -26.943$\pm$0.076 & 0.214$\pm$0.042 & 14.62 & 12.69 & 13.25 & 0  \\
Hivel5 & 212.4778050 & 33.71299000 & 34 & F6 & -252$\pm$10 & -17.612$\pm$0.019 & -16.573$\pm$0.034 & 0.105$\pm$0.019 & 13.09 & 12.51 & 13.51 & 1  \\
Hivel6 & 240.5153620 & 9.52531900 & 153 & G2 & 172$\pm$8 & -3.737$\pm$0.035 & -24.323$\pm$0.023 & 0.126$\pm$0.024 & 13.63 & 12.35 & 12.54 & 1  \\
Hivel7 & 214.9762830 & 37.66936600 & 31 & F0 & -245$\pm$15 & -42.914$\pm$0.038 & -17.258$\pm$0.050 & 0.222$\pm$0.035 & 15.83 & 19.25 & 19.17 & 1  \\
Hivel8 & 194.7713607 & -2.54662930 & 127 & G7 & 92$\pm$5 & 26.920$\pm$0.140 & -134.331$\pm$0.068 & 0.723$\pm$0.063 & 13.61 & 13.07 & 14.00 & 0  \\
Hivel9 & 231.8486900 & 36.03446500 & 174 & G5 & -89$\pm$4 & -15.238$\pm$0.027 & -14.224$\pm$0.041 & 0.103$\pm$0.020 & 11.19 & 10.37 & 11.98 & 0  \\
Hivel10 & 329.7058060 & 1.35603100 & 24 & F7 & -52$\pm$12 & -21.883$\pm$0.068 & -30.680$\pm$0.056 & 0.213$\pm$0.036 & 14.03 & 12.97 & 13.95 & 0  \\
Hivel11 & 190.6504160 & 52.56223400 & 72 & G6 & 99$\pm$6 & -12.198$\pm$0.022 & -13.476$\pm$0.022 & 0.092$\pm$0.015 & 13.42 & 15.90 & 18.79 & 1  \\
Hivel12 & 187.4196460 & 24.50598100 & 153 & G3 & 241$\pm$6 & -15.178$\pm$0.035 & -15.656$\pm$0.026 & 0.111$\pm$0.022 & 13.37 & 12.75 & 13.84 & 1  \\
Hivel13 & 256.3047540 & 19.94524300 & 65 & G3 & -146$\pm$7 & -8.992$\pm$0.021 & -26.154$\pm$0.024 & 0.146$\pm$0.017 & 13.91 & 15.39 & 17.83 & 1  \\
Hivel14 & 288.3732000 & 42.08276000 & 84 & G7 & -316$\pm$10 & -4.459$\pm$0.031 & 12.781$\pm$0.031 & 0.092$\pm$0.017 & 14.39 & 16.17 & 16.80 & 0  \\
Hivel15 & 250.9465099 & 43.60750040 & 59 & F5 & -70$\pm$13 & -16.454$\pm$0.070 & -36.536$\pm$0.106 & 0.231$\pm$0.042 & 16.03 & 15.51 & 18.34 & 1  \\
Hivel16 & 216.9762060 & 29.84803200 & 115 & F0 & -40$\pm$9 & -10.328$\pm$0.034 & -28.546$\pm$0.037 & 0.159$\pm$0.021 & 13.38 & 12.96 & 13.64 & 1  \\
Hivel17 & 244.1838800 & 17.86343200 & 57 & F0 & 23$\pm$16 & -49.937$\pm$0.052 & -19.192$\pm$0.049 & 0.291$\pm$0.049 & 15.54 & 12.51 & 13.81 & 1  \\
Hivel18 & 258.1346050 & 40.47350100 & 90 & G3 & -218$\pm$11 & -20.574$\pm$0.035 & 4.683$\pm$0.040 & 0.123$\pm$0.021 & 12.68 & 14.89 & 18.18 & 1  \\
Hivel19 & 182.5150380 & 0.98761100 & 91 & G7 & 223$\pm$7 & -22.117$\pm$0.055 & -30.106$\pm$0.036 & 0.195$\pm$0.033 & 13.08 & 13.61 & 14.68 & 1  \\
Hivel20 & 207.1676590 & 52.84460600 & 57 & F2 & -103$\pm$15 & -24.004$\pm$0.036 & -24.524$\pm$0.034 & 0.179$\pm$0.025 & 15.23 & 12.36 & 13.51 & 1  \\
\enddata
\tablecomments{
The 4th to 21st columns of Table~\ref{tab:general_catalog} are listed here for the 20 fastest HiVelSCs, and the measurement values and
the uncertainties are shown in the same column in this table but separated into two columns in Table~\ref{tab:general_catalog}.
$^{a}$ r band signal to noise ratio from LAMOST;
$^{b}$ Spectral type given by the LAMOST 1D pipeline;
$^{c}$ $RV$ given by the LASP, which is corrected a radial velocity zero-point of $-$5.4 km~s$^{-1}$ mentioned in
the last two paragraphs of Section~\ref{sec:sample};
$^{d}$ Proper motions from Gaia;
$^{e}$ Parallax from Gaia;
$^{f}$ $G$, $G_{\rm BP}$ and $G_{\rm RP}$ magnitudes from Gaia;
$^{g}$ A flag to show whether a candidate has more conservative astrometric parameters.
}
\end{deluxetable*}
\end{longrotatetable}

\begin{longrotatetable}
\begin{deluxetable*}{lllllllllllll}
\tablecaption{Spatial Positions and Velocities of the 20 fastest HiVelSCs \label{tab:kinematic}}
\tablewidth{0pt}
\tablehead{
\colhead{ID} & \colhead{$x$$^{a}$} & \colhead{$y$$^{a}$} & \colhead{$z$$^{a}$} & \colhead{$r_{\rm GC}$$^{b}$} & \colhead{$V_{\rm x}$$^{c}$} & \colhead{$V_{\rm y}$$^{c}$}
& \colhead{$V_{\rm z}$$^{c}$} & \colhead{$V_{\rm GC}$$^{d}$} & \colhead{$e^{e}$} & \colhead{$Z_{\rm max}$$^{f}$} & \colhead{$r_{\rm min}$$^{g}$} & \colhead{E-$\Phi$($\infty$)$^{h}$} \\
\colhead{} & \colhead{(kpc)} & \colhead{(kpc)} & \colhead{(kpc)} & \colhead{(kpc)} & \colhead{(km s$^{-1}$)} & \colhead{(km s$^{-1}$)} & \colhead{(km s$^{-1}$)} &
\colhead{(km s$^{-1}$)} & \colhead{} & \colhead{(kpc)} & \colhead{(kpc)} & \colhead{(kpc$^{2}$ Myr$^{-2}$)}
}
\startdata
Hivel1 & -5.9$_{-0.3}^{+0.4}$ & 5.1$_{-0.6}^{+0.8}$ & 6.4$_{-0.8}^{+1.0}$ & 10.1$_{-0.6}^{+0.9}$ & -88$_{-8}^{+7}$ & -717$_{-127}^{+104}$ & 573$_{-85}^{+108}$ & 922$_{-136}^{+168}$ &
$-$ & $-$ & 13.7$_{-1.4}^{+1.9}$ & 0.0033$_{-0.0013}^{+0.0019}$  \\
Hivel2 & -9.4$_{-0.2}^{+0.2}$ & 2.0$_{-0.3}^{+0.4}$ & 4.5$_{-0.6}^{+0.8}$ & 10.6$_{-0.4}^{+0.6}$ & 56$_{-5}^{+4}$ & -850$_{-182}^{+132}$ & 256$_{-60}^{+85}$ & 888$_{-142}^{+197}$ &
$-$ & $-$ & 19.1$_{-1.4}^{+1.8}$ & 0.0030$_{-0.0012}^{+0.0020}$  \\
Hivel3 & -8.9$_{-0.0}^{+0.0}$ & 0.1$_{-0.0}^{+0.0}$ & 0.3$_{-0.0}^{+0.0}$ & 8.9$_{-0.0}^{+0.1}$ & -78$_{-14}^{+12}$ & -866$_{-80}^{+72}$ & 110$_{-8}^{+10}$ & 876$_{-73}^{+81}$ &         $-$ & $-$ & 8.9$_{-0.1}^{+0.1}$ & 0.0028$_{-0.0007}^{+0.0007}$  \\
Hivel4 & -10.0$_{-0.4}^{+0.3}$ & 0.5$_{-0.1}^{+0.1}$ & 4.4$_{-0.7}^{+1.0}$ & 10.9$_{-0.5}^{+0.8}$ & -613$_{-137}^{+95}$ & -621$_{-206}^{+142}$ & -41$_{-36}^{+24}$ & 874$_{-168}^{+247}$ & $-$ & $-$ & 92.9$_{-54.6}^{+184.4}$ & 0.0026$_{-0.0012}^{+0.0026}$  \\
Hivel5 & -6.7$_{-0.2}^{+0.4}$ & 2.5$_{-0.4}^{+0.6}$ & 9.0$_{-1.3}^{+2.0}$ & 11.5$_{-0.9}^{+1.6}$ & -73$_{-11}^{+7}$ & -857$_{-237}^{+145}$ & 66$_{-43}^{+70}$ & 862$_{-147}^{+242}$ &
$-$ & $-$ & 74.4$_{-6.4}^{+7.4}$ & 0.0028$_{-0.0013}^{+0.0025}$  \\
Hivel6 & -2.6$_{-0.9}^{+1.2}$ & 2.1$_{-0.4}^{+0.5}$ & 5.3$_{-0.9}^{+1.2}$ & 6.3$_{-0.4}^{+0.8}$ & 639$_{-84}^{+113}$ & -454$_{-168}^{+126}$ & -109$_{-51}^{+39}$ & 791$_{-142}^{+199}$ & $-$ & $-$ & 195.7$_{-105.9}^{+541.8}$ & 0.0018$_{-0.0010}^{+0.0018}$  \\
Hivel7 & -7.6$_{-0.1}^{+0.1}$ & 1.5$_{-0.2}^{+0.3}$ & 4.2$_{-0.6}^{+0.8}$ & 8.8$_{-0.2}^{+0.4}$ & -382$_{-73}^{+48}$ & -676$_{-168}^{+112}$ & 138$_{-50}^{+73}$ & 789$_{-126}^{+194}$ & $-$ & $-$ & 18.2$_{-1.9}^{+2.2}$ & 0.0020$_{-0.0010}^{+0.0016}$  \\
Hivel8 & -7.8$_{-0.0}^{+0.0}$ & -0.6$_{-0.1}^{+0.0}$ & 1.2$_{-0.1}^{+0.1}$ & 7.9$_{-0.0}^{+0.0}$ & 606$_{-44}^{+51}$ & -334$_{-51}^{+43}$ & -357$_{-40}^{+35}$ & 778$_{-69}^{+81}$ &
$-$ & $-$ & $-$ & 0.0018$_{-0.0005}^{+0.0007}$  \\
Hivel9 & -5.3$_{-0.5}^{+0.7}$ & 4.6$_{-0.8}^{+1.1}$ & 8.0$_{-1.3}^{+2.0}$ & 10.6$_{-1.0}^{+1.7}$ & 172$_{-31}^{+46}$ & -631$_{-205}^{+139}$ & 349$_{-68}^{+102}$ & 741$_{-156}^{+236}$ & $-$ & $-$ & 20.1$_{-3.4}^{+4.6}$ & 0.0018$_{-0.0012}^{+0.0019}$  \\
Hivel10 & -6.4$_{-0.3}^{+0.4}$ & 3.1$_{-0.4}^{+0.7}$ & -2.9$_{-0.6}^{+0.4}$ & 7.7$_{-0.1}^{+0.3}$ & 705$_{-102}^{+154}$ & -214$_{-96}^{+61}$ & 16$_{-10}^{+8}$ & 737$_{-114}^{+174}$ &
$-$ & $-$ & 176.9$_{-65.1}^{+177.2}$ & 0.0015$_{-0.0008}^{+0.0015}$  \\
Hivel11 & -10.9$_{-0.6}^{+0.4}$ & 3.8$_{-0.5}^{+0.8}$ & 9.8$_{-1.4}^{+2.0}$ & 15.2$_{-1.2}^{+2.0}$ & -221$_{-43}^{+28}$ & -583$_{-181}^{+119}$ & 371$_{-37}^{+58}$ & 727$_{-123}^{+186}$ & $-$ & $-$ & 19.0$_{-4.0}^{+5.8}$ & 0.0018$_{-0.0010}^{+0.0015}$  \\
Hivel12 & -8.6$_{-0.1}^{+0.1}$ & -0.8$_{-0.2}^{+0.1}$ & 9.0$_{-1.4}^{+2.3}$ & 12.5$_{-1.0}^{+1.8}$ & -215$_{-56}^{+35}$ & -675$_{-232}^{+144}$ & 160$_{-23}^{+15}$ & 726$_{-138}^{+230}$ & $-$ & $-$ & 27.7$_{-10.3}^{+24.4}$ & 0.0016$_{-0.0010}^{+0.0018}$  \\
Hivel13 & -3.8$_{-0.4}^{+0.6}$ & 3.7$_{-0.4}^{+0.5}$ & 3.6$_{-0.4}^{+0.5}$ & 6.4$_{-0.1}^{+0.3}$ & 508$_{-59}^{+79}$ & -503$_{-89}^{+69}$ & -87$_{-4}^{+4}$ & 718$_{-87}^{+121}$ &
$-$ & $-$ & $-$ & 0.0013$_{-0.0007}^{+0.0010}$  \\
Hivel14 & -5.2$_{-0.5}^{+0.7}$ & 10.1$_{-1.6}^{+2.4}$ & 2.6$_{-0.4}^{+0.6}$ & 11.6$_{-1.2}^{+2.0}$ & -593$_{-120}^{+82}$ & -9$_{-11}^{+13}$ & 395$_{-72}^{+109}$ & 713$_{-108}^{+160}$ & $-$ & $-$ & 10.1$_{-1.3}^{+2.3}$ & 0.0014$_{-0.0008}^{+0.0015}$  \\
Hivel15 & -7.0$_{-0.2}^{+0.3}$ & 3.0$_{-0.5}^{+0.7}$ & 2.9$_{-0.4}^{+0.7}$ & 8.1$_{-0.1}^{+0.3}$ & 614$_{-94}^{+147}$ & -277$_{-116}^{+73}$ & 212$_{-39}^{+60}$ & 705$_{-121}^{+191}$ & $-$ & $-$ & 16.1$_{-1.4}^{+2.0}$ & 0.0013$_{-0.0008}^{+0.0014}$  \\
Hivel16 & -6.6$_{-0.2}^{+0.2}$ & 1.6$_{-0.2}^{+0.2}$ & 5.9$_{-0.7}^{+0.9}$ & 9.0$_{-0.3}^{+0.5}$ & 397$_{-46}^{+58}$ & -566$_{-119}^{+94}$ & 90$_{-17}^{+20}$ & 698$_{-104}^{+132}$ &
$-$ & $-$ & 37.3$_{-5.8}^{+8.3}$ & 0.0013$_{-0.0007}^{+0.0012}$  \\
Hivel17 & -6.0$_{-0.4}^{+0.4}$ & 1.4$_{-0.2}^{+0.3}$ & 2.3$_{-0.4}^{+0.4}$ & 6.6$_{-0.1}^{+0.2}$ & -21$_{-13}^{+13}$ & -474$_{-143}^{+117}$ & 512$_{-80}^{+95}$ & 697$_{-137}^{+165}$ & $-$ & $-$ & 6.8$_{-0.0}^{+0.1}$ & 0.0010$_{-0.0007}^{+0.0013}$  \\
Hivel18 & -5.4$_{-0.4}^{+0.6}$ & 6.0$_{-0.8}^{+1.3}$ & 4.7$_{-0.6}^{+1.0}$ & 9.3$_{-0.6}^{+1.1}$ & -328$_{-52}^{+38}$ & -304$_{-76}^{+57}$ & 538$_{-92}^{+139}$ & 697$_{-110}^{+170}$ & $-$ & $-$ & 8.8$_{-1.0}^{+1.6}$ & 0.0013$_{-0.0008}^{+0.0014}$  \\
Hivel19 & -7.7$_{-0.1}^{+0.1}$ & -2.4$_{-0.5}^{+0.3}$ & 4.6$_{-0.7}^{+1.0}$ & 9.3$_{-0.3}^{+0.6}$ & -81$_{-25}^{+17}$ & -659$_{-177}^{+118}$ & -207$_{-87}^{+62}$ & 696$_{-131}^{+198}$ & $-$ & $-$ & 195.7$_{-186.0}^{+283.0}$ & 0.0012$_{-0.0008}^{+0.0015}$  \\
Hivel20 & -8.8$_{-0.1}^{+0.1}$ & 2.5$_{-0.3}^{+0.4}$ & 5.0$_{-0.6}^{+0.8}$ & 10.5$_{-0.4}^{+0.6}$ & -31$_{-9}^{+8}$ & -609$_{-136}^{+103}$ & 324$_{-55}^{+69}$ & 691$_{-118}^{+149}$ &
$-$ & $-$ & 13.8$_{-1.2}^{+1.9}$ & 0.0012$_{-0.0007}^{+0.0012}$  \\
\enddata
\tablecomments{
The 26th to 61st columns of Table~\ref{tab:general_catalog} are listed here for the 20 fastest HiVelSCs, which includes spatial positions and velocities. The median value,
the lower uncertainty, and the upper uncertainty of each parameter are shown in the same column in this table but separated into three columns in Table~\ref{tab:general_catalog} (from 26th to 61st columns).
$^{a}$ Galactocentric spatial position ($x$, $y$, $z$);
$^{b}$ Galactocentric distance;
$^{c}$ Galactocentric spatial velocity ($V_{\rm x}$, $V_{\rm y}$, $V_{\rm z}$);
$^{d}$ Galactocentric total velocity;
$^{e}$ Orbit eccentricity e, ``$-$'' represents e is not provided;
$^{f}$ Orbital maximum height above the Galactic disk, ``$-$'' represents Z$_{\rm max}$ is not provided;
$^{g}$ Minimum crossing radius during the orbital trace back;
$^{h}$ Energy difference between orbital energy and potential energy at infinity;
$^{i}$ For a star, if less than 1$\%$ orbits of a Monte-Carlo simulation with N realizations intersect with
the Galactic disk, we do not provide $r_{\rm min}$ and its errors, which is represented by the ``$-$''. \\
Note that, we only provide e and Z$_{\rm max}$ for 246 bound HiVelSCs with Z$_{\rm max} \leq 200$ kpc, and
``$-$'' represents e and Z$_{\rm max}$ are not provided.}
\end{deluxetable*}
\end{longrotatetable}

\begin{figure*}
\begin{center}
\includegraphics[scale=0.45,angle=0]{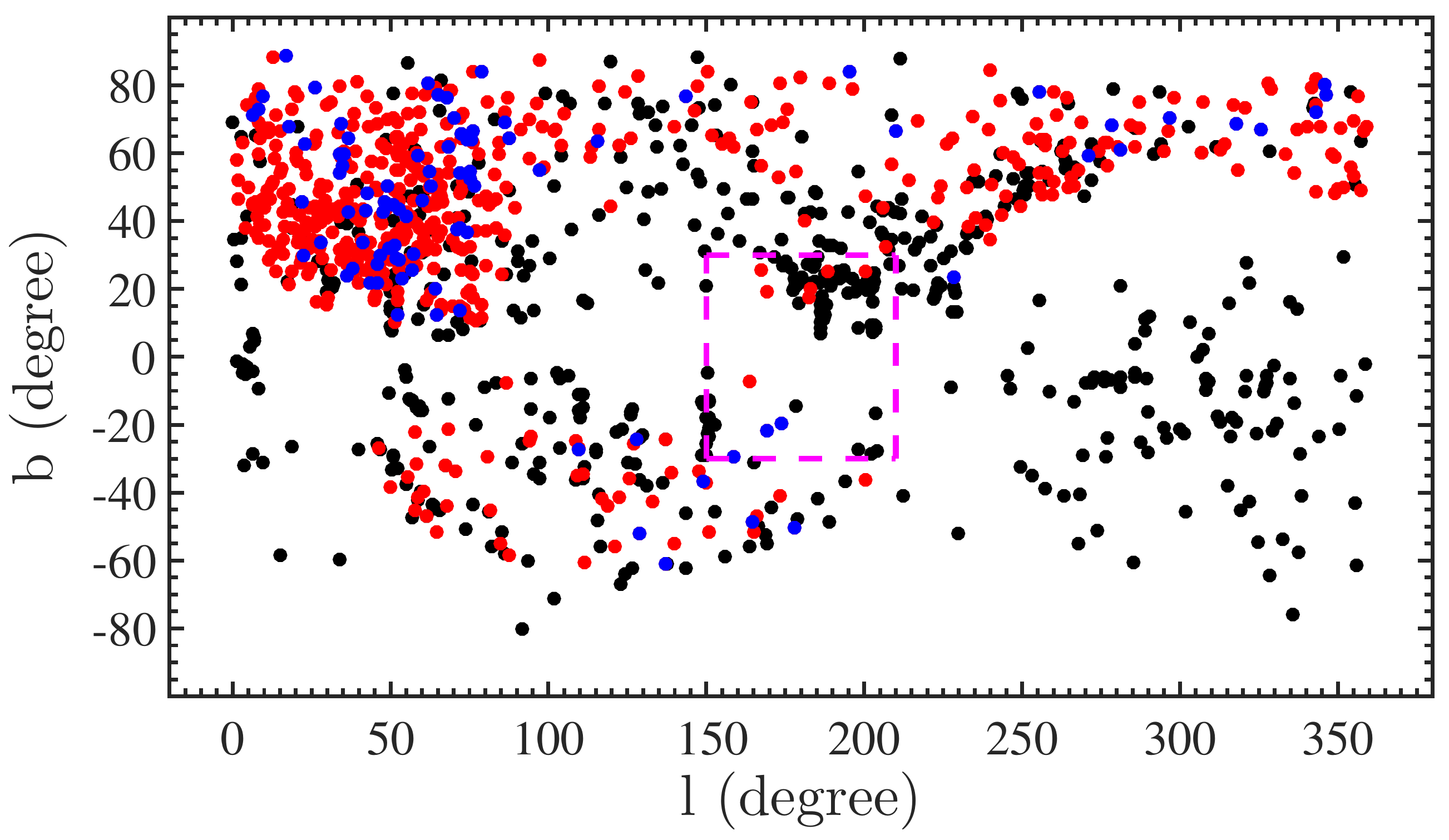}
\caption{The spatial distribution of 591 HiVelSCs in the Galactic coordinates. The black solid dots are 558 stars in the open fast stars catalog,
and the red and blue solid dots represent 591 HiVelSCs, 92 of which are conservative HiVelSCs shown by blue solid dots and selected by more conservative criteria
described in Section~\ref{sec: other_discussion}. \label{fig:gl_gb}}
\end{center}
\end{figure*}

The spatial distribution of the 591 HiVelSCs in Galactic coordinates is plotted in Figure~\ref{fig:gl_gb}, and
the black solid dots are 558 stars in the OFSC. The red and blue solid dots represent 591 HiVelSCs, and the blue
solid dots are 92 conservative HiVelSCs, which are introduced in detail in Section~\ref{sec: other_discussion}.
The dashed magenta rectangle shows the region of LAMOST Galactic anticentre survey \citep{2015MNRAS.448..855Y},
which collects over 3.6 million spectra accounting for at least 34$\%$ of all DR7 released spectra,
but only about 10 HiVelSCs are found in this region, which account for about $1.7\%$ of all HiVelSCs.

\begin{figure*}
\begin{center}
\includegraphics[scale=0.45,angle=0]{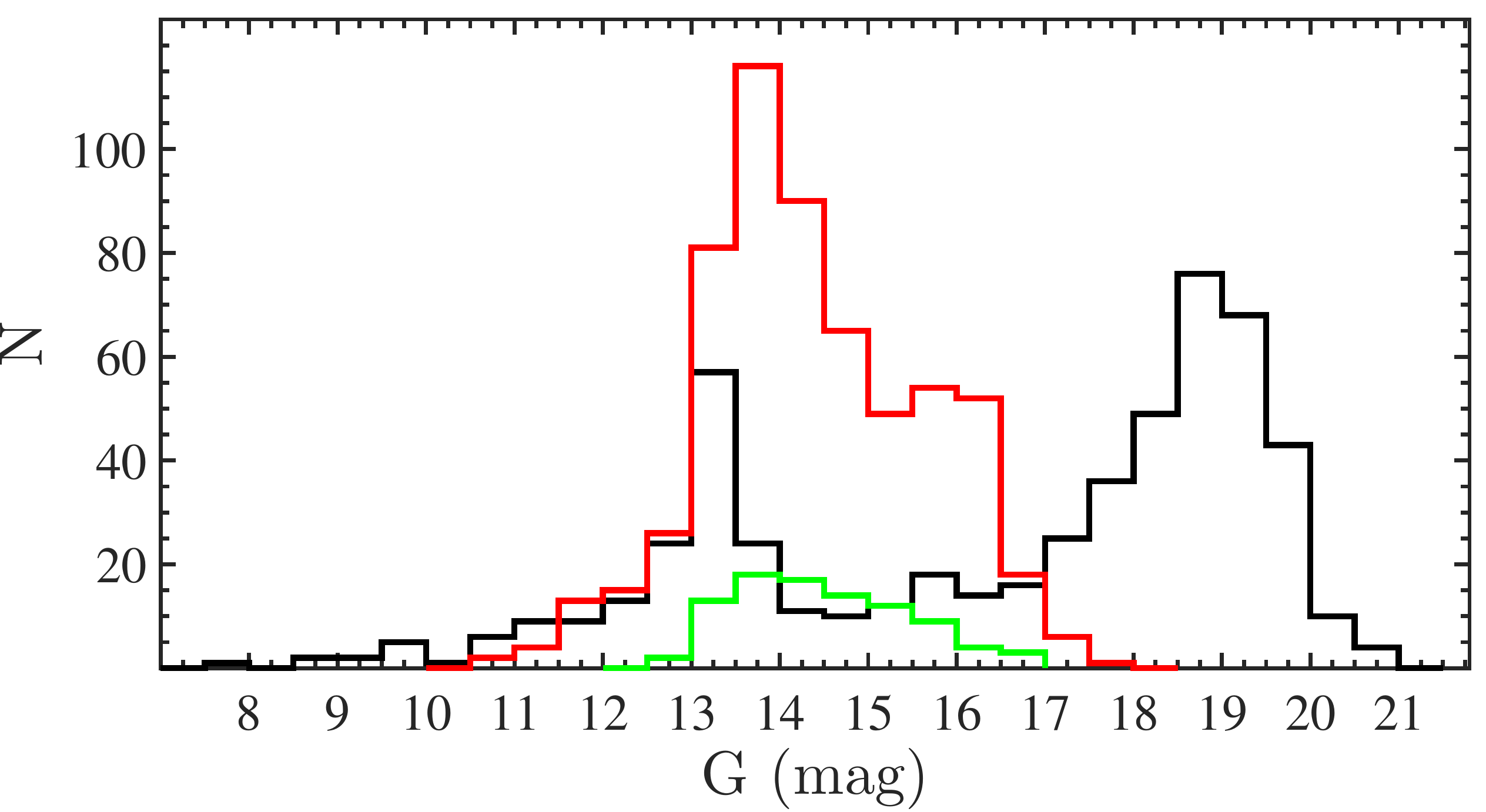}
\caption{$G$ magnitude distribution of Gaia DR2. The black and red histograms are the $G$ magnitude distributions for 558 stars in the open fast stars catalog and
591 HiVelSCs, respectively, and the green one is for 92 conservative HiVelSCs introduced in Section~\ref{sec: other_discussion}. The red and greed distributions
miss the second mode at the faint end of magnitude distribution of OFSC because LAMOST is magnitude limited (SDSS $r$-band magnitude less than 17.8 mag \citep{2018ApJ...863...87D}). \label{fig:Gmag}}
\end{center}
\end{figure*}

The $G$ magnitude distribution of Gaia DR2 is shown in Figure~\ref{fig:Gmag}, and the black histogram is the distribution for 558 stars in the the open fast stars catalog. The red histogram is the distribution for 591 HiVelSCs, and the green one is for 92 conservative HiVelSCs introduced in Section~\ref{sec: other_discussion}. From this figure, we can clearly see that our HiVelSC sample contains relatively more objects in the magnitude range from 13 mag to 16.5 mag, and we miss the second mode at the faint end of magnitude distribution of the open fast stars catalog because LAMOST is magnitude limited.

\begin{figure*}
\begin{center}
\includegraphics[scale=0.4,angle=0]{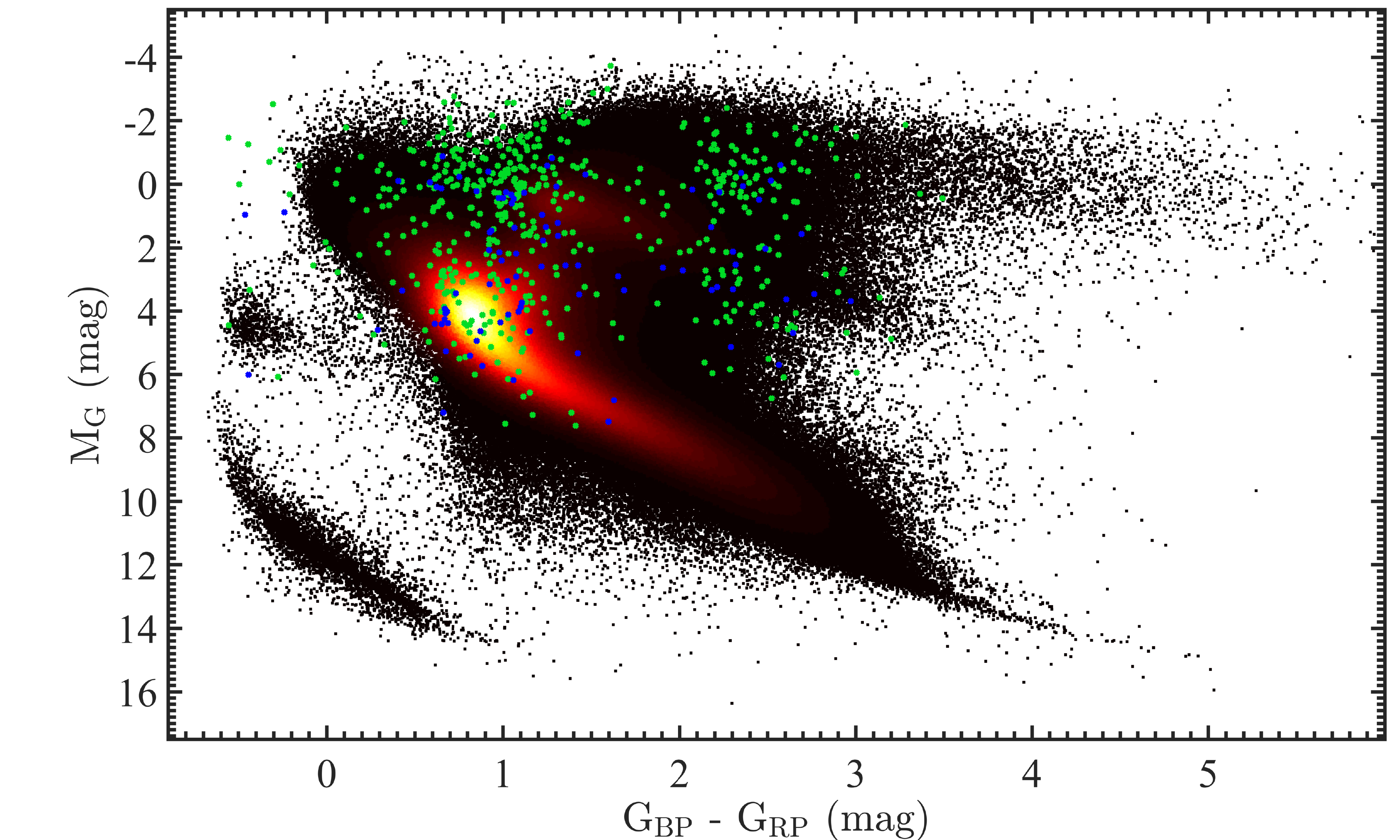}
\caption{The Hertzsprung-Russell diagram for all the 591 HiVelSCs (green and blue x-marks), 92 of which are conservative HiVelSCs (blue x-marks)
introduced in Section~\ref{sec: other_discussion}. The black solid dots represent over 8.48 million ``low-f'' spectra mentioned
in Section~\ref{sec:sample}. \label{fig:cmd}}
\end{center}
\end{figure*}

Figure~\ref{fig:cmd} shows the Hertzsprung-Russell diagram for 591 HiVelSCs (green and blue x-marks) and over 8
million ``low-f'' spectra introduced in Section~\ref{sec:sample} (black solid dots). The x-axis represents the color index in the Gaia Blue
Pass (BP) and Red Pass (RP) bands (BP-RP), and the y-axis gives the absolute magnitude of Gaia G band.
It should be noted that we do not consider extinction when construct the HR diagram, because of the caveats
using the extinction in the G band for individual sources \citep{2018A&A...616A...8A}. From this figure, we
can see that the majority of 591 HiVelSCs are giant stars.

As mentioned in Section~\ref{sec:sample}, seven Galactic potential models
are used to estimate escape velocities for 591 HiVelSCs, and Table~\ref{tab:vesc} lists these escape velocities for the 20 fastest HiVelSCs. In this table, $V_{\rm esc}$(W), $V_{\rm esc}$(G), $V_{\rm esc}$(Ke), $V_{\rm esc}$(Ko), $V_{\rm esc}$(P), $V_{\rm esc}$(M), and $V_{\rm esc}$(X) are escape velocities estimated by using the seven potential models of Watkins+2019, Gnedin+2005, Kenyon+2014, Koposov+2010, Paczynski+1990, MWPotential2014, and Xue+2008, respectively.
Meanwhile, we estimate the probability $P_{\rm ub}$ of being unbound from the Milky Way, which is defined as the percentage of $V_{\rm GC}$ larger than
escape velocity in 1000 MC realizations of $r_{\rm GC}$ and $V_{\rm GC}$, and the Table~\ref{tab:unbound} lists these probabilities for the 20 fastest HiVelSCs. In this table, $P_{\rm ub}$(W), $P_{\rm ub}$(G), $P_{\rm ub}$(Ke), $P_{\rm ub}$(Ko), $P_{\rm ub}$(P), $P_{\rm ub}$(M), and $P_{\rm ub}$(X) are unbound probabilities estimated by using the above seven potential models. Table~\ref{tab:punbound_50} lists the number of HiVelSCs with $P_{\rm ub} > 50\%$ under each potential model, and we can see
that at least 43 HiVelSCs have $P_{\rm ub} > 50\%$ of being unbound in Watkins+2019 potential model, and at most 304 HiVelSCs have $P_{\rm ub} > 50\%$
of being unbound in Xue+2008 potential model, and the amount of unbound HiVelSCs with $P_{\rm ub} > 50\%$ in other potential models is given by a
number that is within the range (43, 304), i.e., larger than 43 and less than 304.

\begin{deluxetable*}{cccccccc}
\tablecaption{The number of HiVelSCs with probabilities of being unbound larger than 50$\%$ ($P_{\rm ub} \geq 50\%$) in each of the seven potential models\label{tab:punbound_50}}
\tablewidth{0pt}
\tablehead{
\colhead{} & \colhead{Watkins+2019} & \colhead{Gnedin+2005} & \colhead{Kenyon+2014} & \colhead{Koposov+2010} & \colhead{Paczynski+1990} & \colhead{MWPotential2014} & \colhead{Xue+2008} \\
}
\startdata
$P_{\rm ub} \geq 50\%$ & 43 & 52 & 59 & 97 & 149 & 211 & 304 \\
\enddata
\end{deluxetable*}

\begin{figure*}
\begin{center}
\includegraphics[scale=0.4,angle=0]{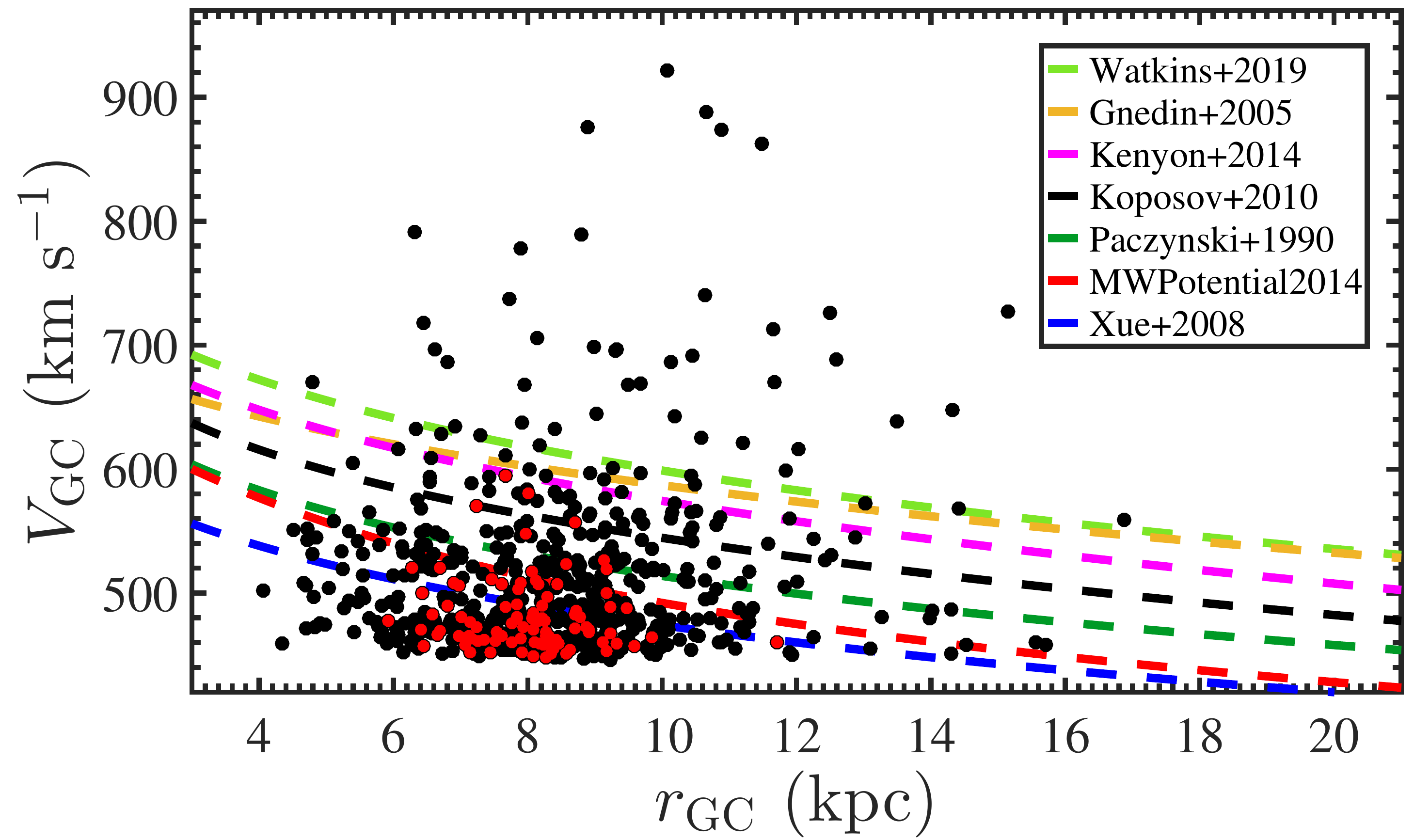}
\caption{Total velocities in the Galactic rest-frame ($V_{\rm GC}$) as a function of Galactocentric distances ($r_{\rm GC})$ for 591 HiVelSCs. The seven dashed curves in different colors are escape velocities at different
distances $r_{\rm GC}$ determined by seven Galactic potential models
\citep{1990ApJ...348..485P, 2005ApJ...634..344G, 2008ApJ...684.1143X, 2010ApJ...712..260K, 2014ApJ...793..122K, 2015ApJS..216...29B, 2019ApJ...873..118W},
and the black and red solid dots are 591 HiVelSCs, 92 of which are conservative HiVelSCs (red solid dots) introduced in Section~\ref{sec: other_discussion}.
\label{fig:r_vgal}}
\end{center}
\end{figure*}

To highlight visually possibly unbound HiVelSCs, we plot the total velocities in the Galactic rest frame
$V_{\rm GC}$ as a function of Galactocentric distances $r_{\rm GC}$ for 591 HiVelSCs (black and red solid dots)
in Figure~\ref{fig:r_vgal}, and also plot escape velocities at different $r_{\rm GC}$ with color short
dashed lines based on the above seven potential models. Considering the total velocity uncertainties, there are at least 43 stars
in Figure~\ref{fig:r_vgal} that can escape from the Galaxy in the most conservative Watkins+2019 potential model (resulting in the largest escaping velocities) with $P_{\rm ub} > 50\%$, and 287 stars are bound in the least conservative potential model of Xue+2008 (resulting in the smallest escaping velocities) with $P_{\rm ub} \leq 50\%$. Whether the remaining 261 stars are marginally unbound with $P_{\rm ub} > 50\%$ or not depends on the potential model used.

In order to mark HiVelSCs with high quality Gaia astrometry, here we adopted equations (C.1), (C.2), and the latter
two criteria of selection A in Appendix C of \citet{2018A&A...616A...2L}, and the equality cuts from (i) to (iv)
in Section 4 of \citet{2019MNRAS.490..157M}. The ``astrometric\_flag'' column in Table~\ref{tab:basic} is used to
determine the quality of astrometric parameters, and the value of ``1'' represents a HiVelSC that satisfies
all above criteria. Using the ``astrometric\_flag'' column, 476 HiVelSCs have conservative high quality astrometric
parameters of equatorial coordinates, parallaxes, and proper motions.

\section{Properties of High Velocity Star Candidates} \label{sec:hvsc}
\subsection{Distribution of Spatial Positions and Velocities} \label{sec:spatial_position_velocity}

\begin{figure*}
\begin{center}
\includegraphics[scale=0.4,angle=0]{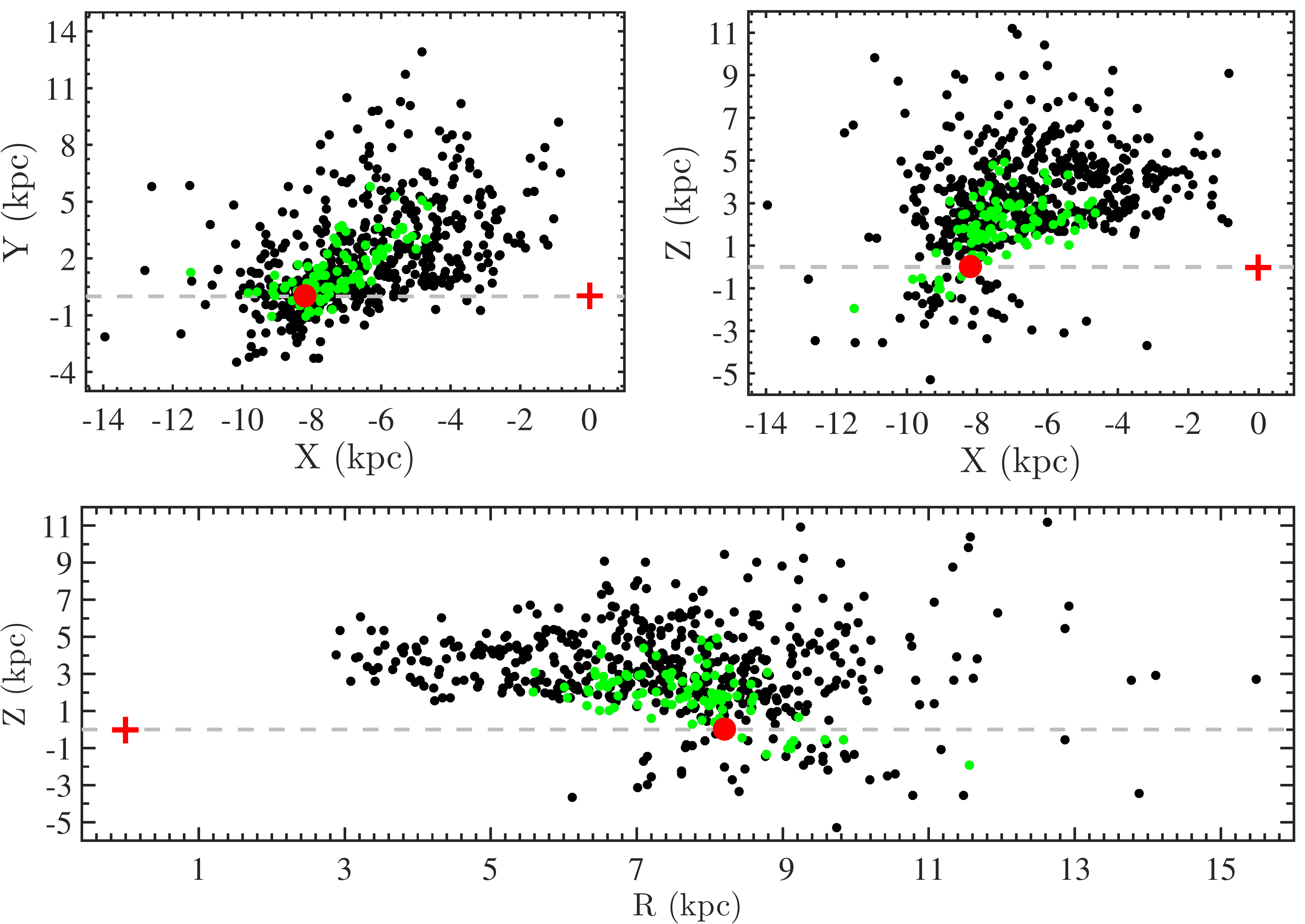}
\caption{The spatial distribution of 591 HiVelSCs. Upper left panel: the distribution on the Galactic plane
. Upper right panel: the distribution in the (X, Z) plane. Bottom panel: the distribution in the Galactocentric
cylindrical coordinates (R, Z). The Sun and GC are marked by a red solid circle and a
red plus, respectively. The black and green solid dots represent 591 HiVelSCs, and the green solid dots are 92 conservative HiVelSCs
introduced in Section~\ref{sec: other_discussion}. \label{fig:XYZ}}
\end{center}
\end{figure*}

The spatial distribution of 591 HiVelSCs is shown in Figure~\ref{fig:XYZ}. The upper
panels are the distribution on the Galactic disk and the (X, Z) plane, and the bottom panel is the distribution
in the Galactocentric cylindrical coordinates (R, Z). From the upper panels, we can see that all the 591 HiVelSCs
are located on the side of the sun having negative $X$, whereas majority of them have positive $Y$ ($\sim$$80.7\%$)
and $Z$ ($\sim$$91\%$). From the bottom panel, we note that about half (49$\%$) of our HiVelSCs lie away from
the stellar disk with $|Z| \geq 3$~kpc, which is the edge of the thick disk \citep{2010ApJ...712..692C}.

\begin{figure*}
\begin{center}
\includegraphics[scale=0.4,angle=0]{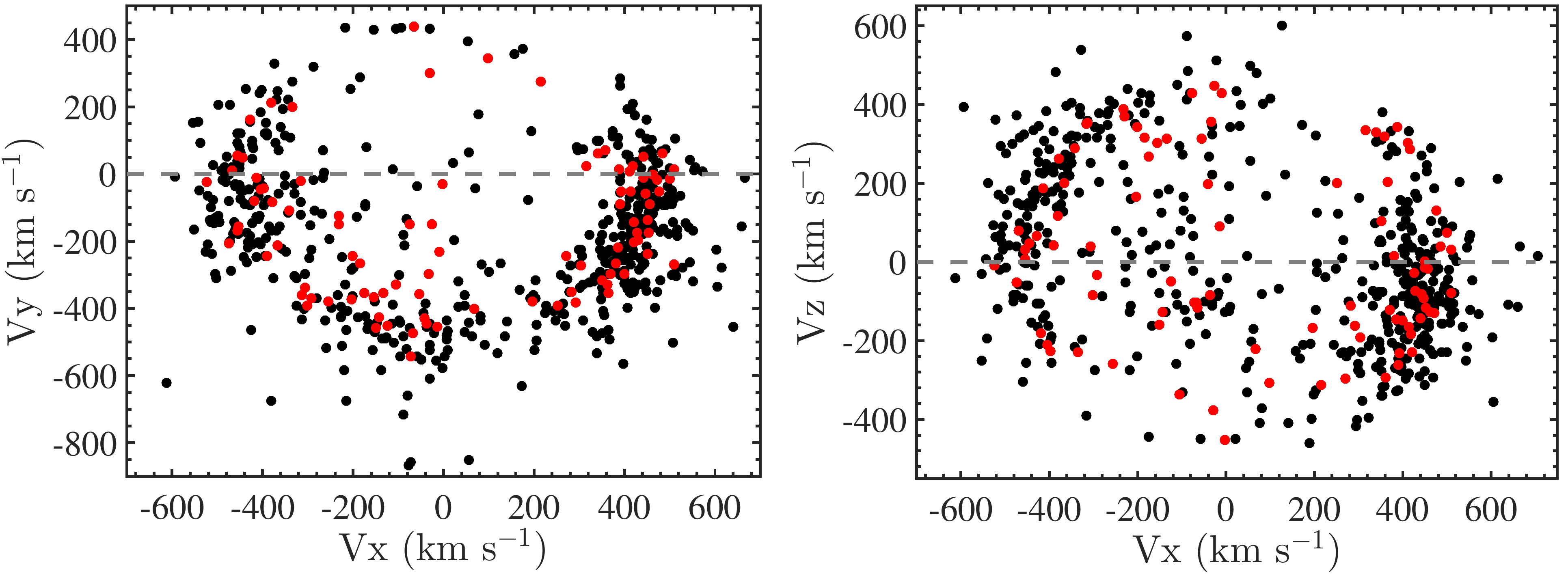}
\caption{Galactocentric velocity distribution of 591 HiVelSCs. The black and red solid dots represent 591 HiVelSCs, and the
red solid dots are 92 conservative HiVelSCs introduced in Section~\ref{sec: other_discussion}.\label{fig:vxvyvz}}
\end{center}
\end{figure*}

Figure~\ref{fig:vxvyvz} shows the Galactocentric velocity components of our HiVelSCs, showing that
they are not clumped in velocity. These stars show symmetry in the
$V_{\rm x}$~-~$V_{\rm z}$ plane but an obvious asymmetry in the $V_{\rm x}$~-~$V_{\rm y}$ plane, and most of
HiVelSCs have negative values of $V_{\rm y}$. There are at most 304 out of 591 HiVelSCs unbound to the Galaxy with
$P_{\rm ub} > 50\%$ and 251 ($\sim$$82.6\%$) of them have negative values of $V_{\rm y}$, assuming the Xue+2008
potential model (The number of unbound ones is smaller if we assume other potential models.) Such an asymmetry
in the $V_{\rm x}$~-~$V_{\rm y}$ plane is consistent with the discussion in
\citet{2019A&A...627A.104D}, and negative $V_{\rm y}$ components (retrograde) tend to be significantly larger than
the positive counterparts (prograde). But this could be a side effect of the large uncertainties in heliocentric distances
\citep{2019A&A...627A.104D, 2011MNRAS.415.3807S}. In order to further exclude stars with large uncertainties on
distances, we use ``astrometric\_flag'' to single out 245 stars with more conservative Gaia astrometry
out of all 304 unbound HiVelSCs mentioned above, and 206 ($\sim$$84\%$) of them still have retrograde $V_{\rm y}$. This result is consistent
with that in \citet{2019A&A...627A.104D}, but the ratio of HiVelSCs with negative values of $V_{\rm y}$ is much higher.
As discussed in literature \citep{2019A&A...627A.104D, 2019MNRAS.490..157M, 2016A&A...588A..41C}, most unbound HiVelSCs having negative
values of $V_{\rm y}$ suggest that a population of high velocity stars that may have an extragalactic provenance.

\subsection{Halo or Disk Stars?} \label{sec:halo_disk}

\begin{figure*}
\begin{center}
\includegraphics[scale=0.4,angle=0]{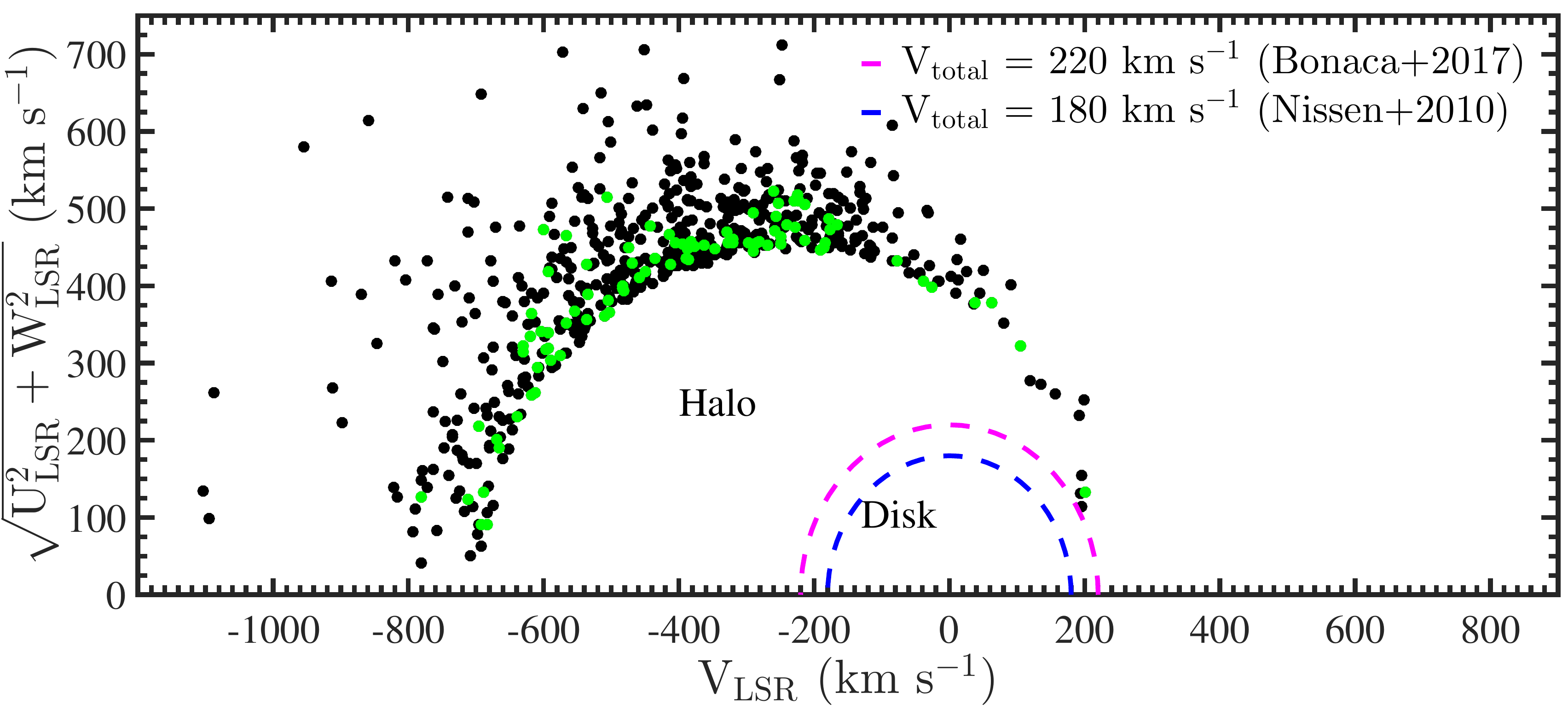}
\caption{Toomre diagram for 591 HiVelSCs, and ($U_{\rm LSR}$, $V_{\rm LSR}$, $W_{\rm LSR}$) is the three dimensional velocity with respect
to the Local Rest Frame. The magenta and blue dashed semicircles represent $V_{\rm total} =  \sqrt{U_{\rm LSR}^2 + V_{\rm LSR}^2 + W_{\rm LSR}^2} = 220$ km s$^{-1}$
and $V_{\rm total} = \sqrt{U_{\rm LSR}^2 + V_{\rm LSR}^2 + W_{\rm LSR}^2} = 180$ km s$^{-1}$, respectively, and they can be used to distinguish halo stars from
disk stars. The black and green solid dots represent 591 HiVelSCs, and the green solid dots are 92 conservative HiVelSCs introduced in Section~\ref{sec: other_discussion}. \label{fig:toomre}}
\end{center}
\end{figure*}

Figure~\ref{fig:toomre} shows the Toomre diagram for 591 HiVelSCs, which has been widely used to
distinguish the thin-disk, thick-disk, and halo stars. On the x axis, we plot the component
$V_{\rm LSR}$ of the space total velocity ($V_{\rm total}$) with respect to the LSR, and on the y axis the perpendicular
component to it, $\sqrt{U_{\rm LSR}^2 + W_{\rm LSR}^2}$. The magenta and blue dashed semicircles are dividing lines
between halo and disk stars, which were defined by \citet{2017ApJ...845..101B} and \citet{2010A&A...511L..10N}, respectively.
The magenta dashed semicircle can be represented by $|V_{\rm total}| \geq 220$ km s$^{-1}$ \citep{2017ApJ...845..101B},
and the blue one can be defined as $|V_{\rm total}| \geq 180$ km s$^{-1}$ \citep{2010A&A...511L..10N}. Here, we adopt the more conservative
magenta dashed semicircle as the dividing line of halo and disk stars. As shown in this figure,
all HiVelSCs are located outside of the magenta dashed semicircle, which means that all the 591 HiVelSCs have halo-like kinematic properties.

Following \citet{2003A&A...410..527B}, we estimate the relative probabilities for the thick-disk-to-halo (TD/H) membership
using equations (1) - (3), which can also be used to distinguish stellar population, and the result shows that relative probabilities of all
591 HiVelSCs of not being halo stars are far less than 0.1 and nearly equal to 0. Such a result indicates that all the 591 HiVelSCs are halo stars with extremely high probabilities, and it is consistent with the result obtained by using the Toomre diagram.

Besides the kinematic methods, chemistry is usually used together with kinematics to study the stellar population and possible origin places, and we use the
method of data-driven Payne (DD-Payne) described in \citet{2019ApJS..245...34X} \citep[see also][]{2017ApJ...849L...9T} to determine $\alpha$ element abundance ([$\alpha$/Fe]) in this work since it was not provided by the LASP. The DD-Payne method also provides atmospheric parameters for 591 HiVelSCs, and these parameters
are included in our ``LAMOSTDR7-GAIADR2-HiVelSC'' catalogue together with LASP atmospheric parameters. Table~\ref{tab:atmospheric} shows these parameters for
the 20 fastest HiVelSCs, where $T_{\rm eff}$\_LASP, log~$g$\_LASP, and [Fe/H]$\_$LASP are atmospheric parameters determined by the LASP,
and $T_{\rm eff}$\_DD-Payne, log~$g$\_DD-Payne, [Fe/H]$\_$DD-Payne and [$\rm \alpha$/Fe]$\_$DD-Payne are parameters obtained by the DD-Payne method.

\begin{figure*}
\begin{center}
\includegraphics[scale=0.5,angle=0]{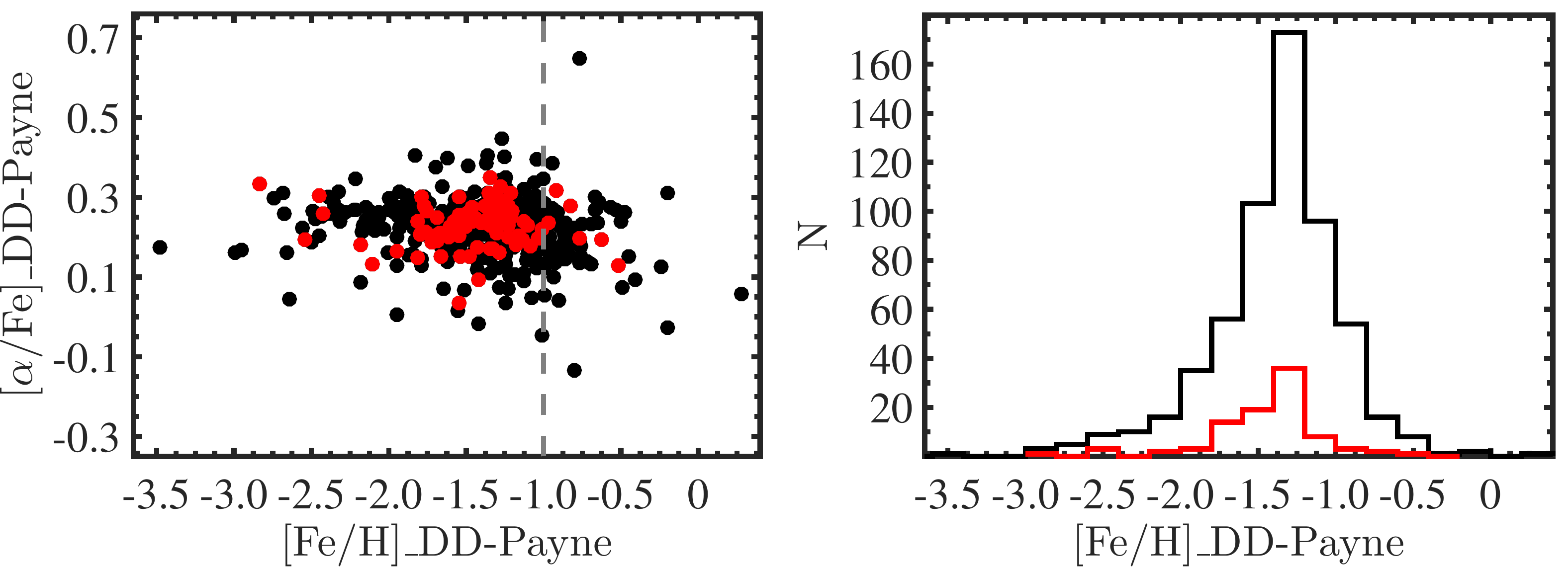}
\caption{Left panel: The distribution of 591 HiVelSCs on the ([$\alpha$/Fe], [Fe/H]) plane. The horizontal and vertical axes are the metallicity ([Fe/H]\_DD-Payne) and the $\alpha$ element abundance ([$\alpha$/Fe]\_DD-Payne), respectively, and they are estimated by the method of data-driven Payne (DD-Payne). Right panel: the distribution of [Fe/H]\_DD-Payne for 591 HiVelSCs, which peaks at about [Fe/H]~$\sim$~$-$1.2 and has a wide range from
near [Fe/H]~$\sim$~$-$3.5 to [Fe/H]~$\sim$~+0.5. The black and red solid dots in the left panel represent 591 HiVelSCs, and the red solid dots are 92
conservative HiVelSCs introduced in Section~\ref{sec: other_discussion}. The black histogram are the distribution of [Fe/H]\_DD-Payne for 591 HiVelSCs, and
the red one is the distribution of [Fe/H]\_DD-Payne for 92 conservative HiVelSCs. \label{fig:alpha_fe}}
\end{center}
\end{figure*}

The left panel of Figure~\ref{fig:alpha_fe} plots the distribution of 591 HiVelSCs on the DD-Payne ($\rm \alpha$, [Fe/H]) plane, and the mean and standard
deviation of the $\alpha$-abundances of our HiVelSCs are about +0.23 dex and 0.07 dex. This is consistent with the results in \citet{2018ApJ...869L..31D} and \citet{2015MNRAS.447.2046H}, which have mean $\alpha$ abundances of $\rm \alpha = +0.22~dex$ and $\rm \alpha = +0.24~dex$, respectively. The right panel
shows the [Fe/H]$\_$DD-Payne distribution of these stars, which peaks at near [Fe/H]~$\sim$~$-$1.2 and has a wide range from near [Fe/H]~$\sim$~$-$3.5 to [Fe/H]~$\sim$~+0.5. From this figure, the 591 kinematically selected halo stars are mostly metal-poor and slightly $\alpha$-enhanced. In addition, their [Fe/H] distribution and its peak value at about [Fe/H]~$\sim$~-1.2 are consistent with the distribution of inner halo stars in \citet{2019ApJ...887..237C} and \citet{2018ApJ...862..163L}. While the mean or peak values of metallicity distributions of inner halo in some research works \citep{2010ApJ...712..692C, 2007Natur.450.1020C, 2015ApJ...809..144X, 2016MNRAS.460.1725D, 2019MNRAS.482.3426M, 2020MNRAS.492.4986Y, 2014A&A...568A...7A, 2013ApJ...763...65A, 2017ApJ...841...59Z} can shift towards more metal poor for several reasons such as bias in halo stars selection as discussed in \citet{2019ApJ...887..237C}, the metallicity distribution of our 591 HiVelSCs is nonetheless broadly consistent with that of the inner halo in these works.

As mentioned in recent works
\citep{2019ApJ...887..237C, 2017ApJ...845..101B, 2018ApJ...863...87D}, a fraction of kinematically defined metal-rich halo stars are in situ components of
the local stellar halo, which display thick-disk chemistry on halo-like orbits and are confined to the range of
$|z| \leq 10$ kpc and $R_{\rm gal} \leq 20$ kpc. Such in situ halo stars are considered to be formed in
the initial collapse of the Milky Way or formed in the disk or bulge and be kinematically heated subsequently
\citep{2019ApJ...887..237C, 2017ApJ...845..101B, 2018ApJ...863...87D}. In our 591 HiVelSCs, 83 ($\sim$14$\%$) of
them are metal-rich halo stars with [Fe/H]$_{\rm DD-Payne} > $-1, and their vertical height and Galactocentric
radius are limited in the range of $|z| \leq 10$ kpc and $R_{\rm gal} \leq 20$ kpc. Thus, they may be the in situ
halo stars \citep{2010A&A...511L..10N, 2019ApJ...887..237C}, and such a lower ratio of about 14$\%$
(but 30$\%$ from \citet{2018ApJ...863...87D} and 50$\%$ from \citet{2017ApJ...845..101B}) is consistent with the result in
\citet{2019ApJ...887..237C} that the bulk of the stellar halo formed from the accretion and tidal disruption.

\begin{figure*}
\begin{center}
\includegraphics[scale=0.4,angle=0]{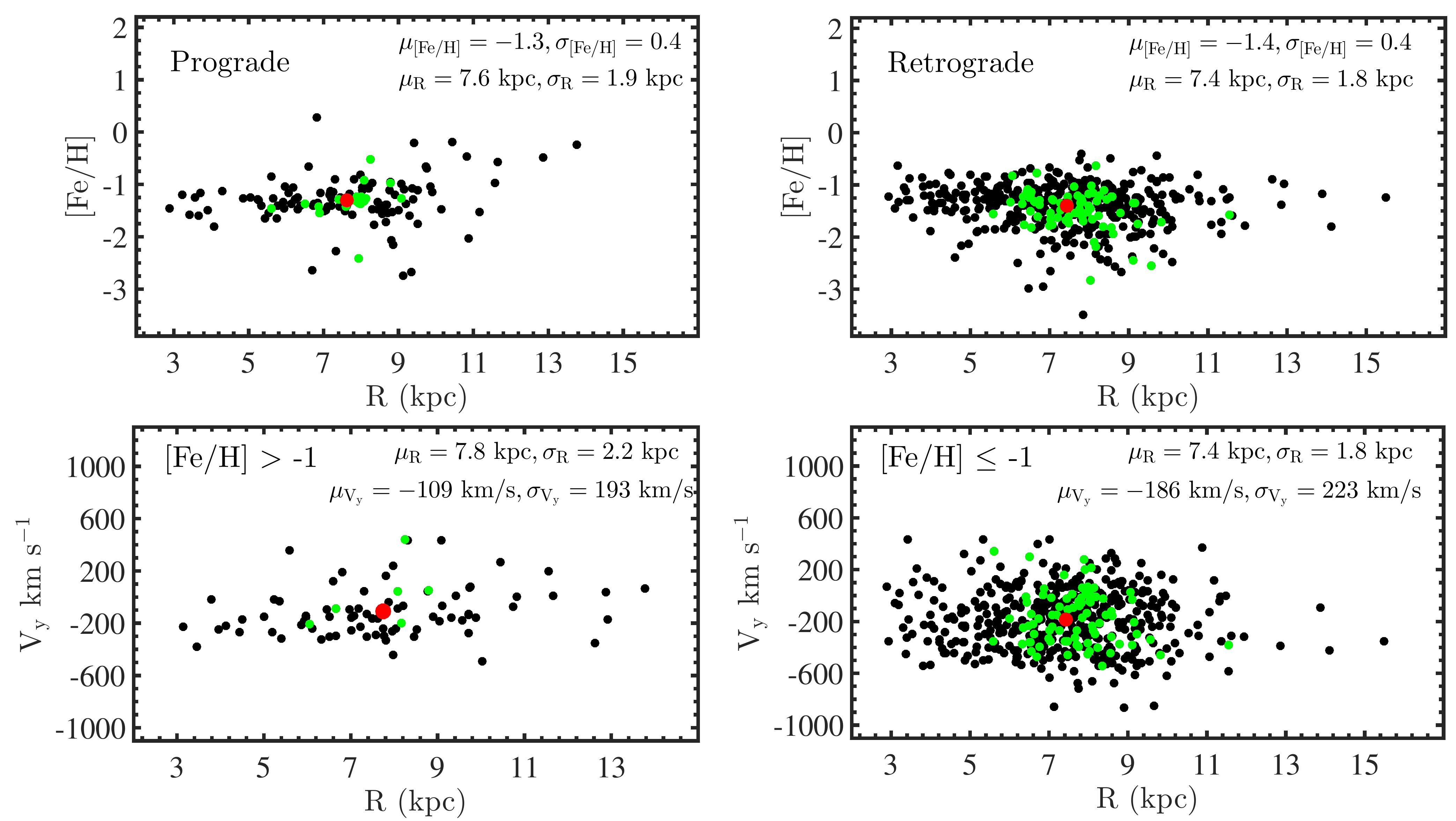}
\caption{Upper panels: Metallicity ([Fe/H]) vs. Galactocentric radius ($R$) separated according to the y-components ($V_{\rm y}$) of the Galactocentric total velocities. The red solid dot marks the mean values of [Fe/H] and $R$. Upper left and right panels are for prograde and retrograde HiVelSCs, respectively. Bottom panels: Distribution of 591 HiVelSCs on the (V$_{\rm y}$, $R$)} plane separated by [Fe/H]. The red solid dot marks the mean values of V$_{\rm y}$ and $R$. Bottom left and right panels are for HiVelSCs with [Fe/H] $>$ $-$1 and [Fe/H] $\leq$ $-$1, respectively. The green solid dots in each panel represent 92 conservative HiVelSCs introduced in Section~\ref{sec: other_discussion}. \label{fig:kinematics_chemistry}
\end{center}
\end{figure*}

In recent studies of the Galactic halo's metallicity distribution \citep{2017ApJ...841...59Z, 2018ApJ...862..163L, 2019ApJ...887..237C}, the majority of the
retrograde stars are more metal-poor than the prograde stars. We thus divide our HiVelSCs into two groups, i.e., prograde and
retrograde HiVelSCs, and plot their distribution on the ([Fe/H], $R$) plane in the upper panels of Figure~\ref{fig:kinematics_chemistry}.
We further divide them into another two types, i.e., metal-rich ([Fe/H] $>$ $-$1) and metal-poor ([Fe/H] $\leq$ $-$1) HiVelSCs, and plotted them on the
($V_{\rm y}$, $R$) plane in the bottom panels. In this figure, we show the mean values of
horizontal and vertical axes as red solid dots.

From the upper panels, we can see that retrograde HiVelSCs have slightly low mean value of [Fe/H] ($\langle$[Fe/H]$\rangle$ = $-$1.4), and this is consistent with the results in the above literature \citep{2017ApJ...841...59Z, 2018ApJ...862..163L, 2019ApJ...887..237C}. In the bottom panels, about 75$\%$ metal-rich HiVelSCs are retrograde, and the ratio for metal-poor HiVelSCs is about 80$\%$. Thus, the majority of our HiVelSCs tend to be retrograde no matter if they are metal-rich or metal-poor. From the bottom panels, we can see that metal-poor HiVelSCs tend to have significantly faster mean retrograde velocities V$_{\rm y}$ ($\langle$V$_{\rm y}$$\rangle$ = $-$186~km s$^{-1}$). Although the results here combining chemistry and kinematics are consistent with recent studies, further verification is needed because the number of stars in our HiVelSCs samples is small and the precisions for chemical and kinematic parameters need further improvement, especially the precisions that correspond to the metallicity and $RV$ from LAMOST low resolution spectra, as well as those of parallaxes and proper motions from Gaia DR2.

\section{Orbital Integration and Origin} \label{sec:orbit}

In order to better understand the ejection locations of our HiVelSCs, we perform numerical orbit integrations and
trace their trajectories back in time using the python package Galpy \citep{2015ApJS..216...29B}. For each star,
weperform 1000 random MC samplings of the equatorial coordinate, $RV$, proper motion, and
parallax considering the measurement errors as discussed in Section~\ref{sec:distance_velocity}, and trace each orbit back to 10 Gyr ago,
with a fixed time-step of 10 Myr, using the MWPotential2014 potential of the Galpy package which consists
of a bulge modeled as a power-law density profile that is exponentially cut-off, a MiyamotoNagaiPotential disk,
and a NFW dark-matter halo \citep{2015ApJS..216...29B}.

We estimate for each HiVelSC the maximum distance above the Galactic plane ($Z_{\rm max}$), the eccentricity ($e$), the minimum
crossing radius ($r_{\rm min}$), and the energy ($E$) of each orbit during the 1000 MC realizations, and
record the Galactocentric coordinates ($x_{\rm c}$, $y_{\rm c}$) at the instant
when $z$ = 0, and define the crossing radius ($r_{\rm c}$) as $r_{\rm c} = \sqrt{x_{\rm c}^{2} + y_{\rm c}^{2}}$
\citep{2019MNRAS.490..157M}. In the case of multiple disk crossing for each orbit, $r_{\rm min}$ is defined as the minimum 
$r_{\rm c}$. In order to determine unbound or bound orbits, we calculate the energy of $E-\Phi(\infty)$, in which $E$
is the orbit energy and $\Phi(\infty)$ is the energy of the convergence potential MWPotential2014 at infinity.

\begin{figure*}
\begin{center}
\includegraphics[scale=0.4,angle=0]{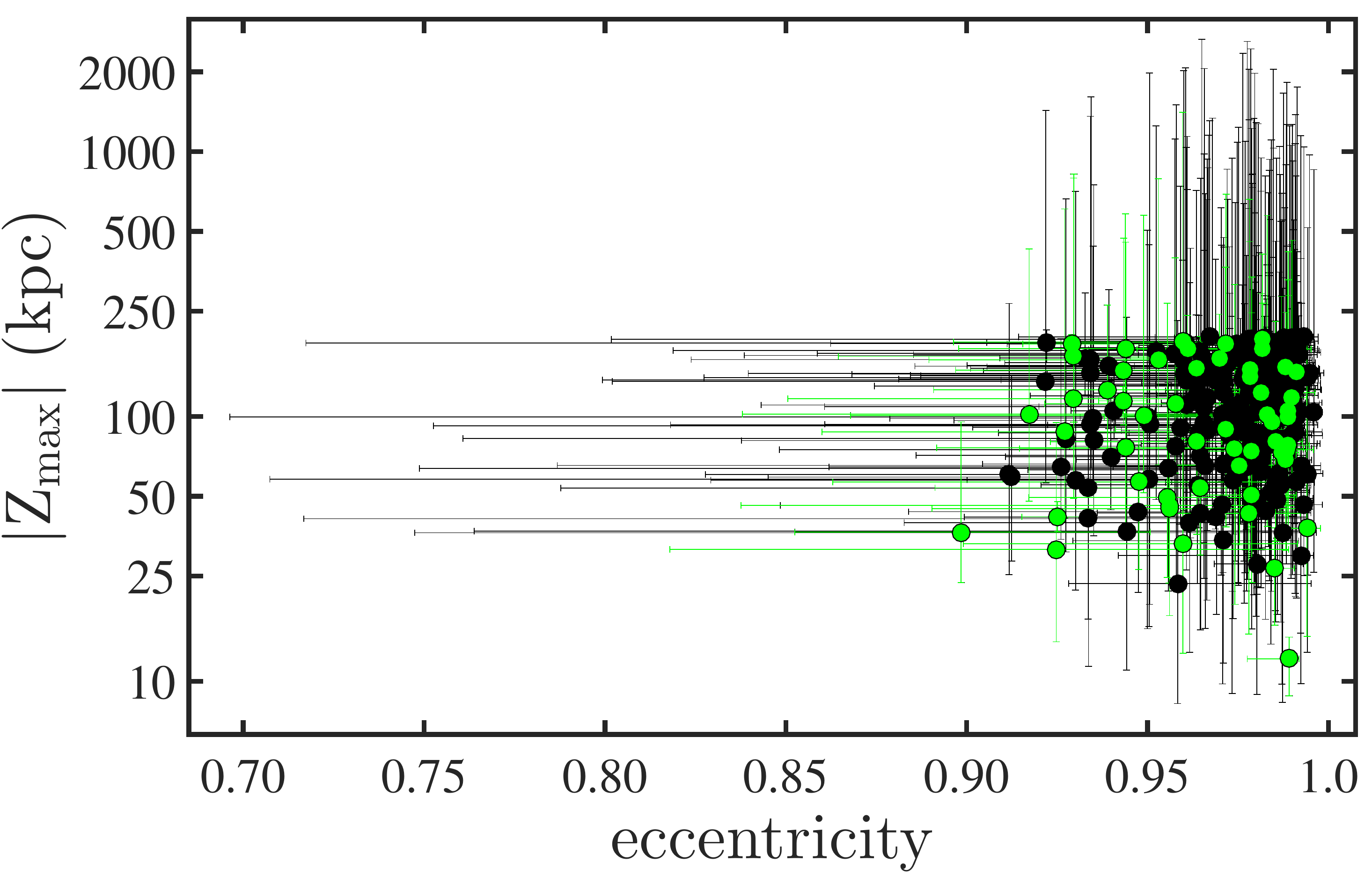}
\caption{The maximum height above the Galactic plane $|Z_{\rm max}|$ as a function of eccentricity $e$, and the y-axis is in log scale. The black solid dots represent 246 bound HiVelSCs with $Z_{\rm max} \leq 200$ kpc, and the green solid dots represent 55 conservative HiVelSCs with $Z_{\rm max} \leq 200$ kpc. The conservative HiVelSCs are introduced in Section~\ref{sec: other_discussion}. \label{fig:zmax_eccentricity}}
\end{center}
\end{figure*}

In Figure~\ref{fig:zmax_eccentricity}, we plot $Z_{\rm max}$ as a function of $e$ only for 246 bound HiVelSCs with $Z_{\rm max} \leq 200$ kpc,
solid dots show the median values of $Z_{\rm max}$ and $e$, and the error bars mark their standard deviations. This plot has the ability
to sort out stars of similar orbits like the Toomre diagram, because eccentricity $e$ represents the shape of
the orbit and $Z_{\rm max}$ represents the amplitude of the orbit vertical oscillation \citep{2013A&A...553A..19B}.
From this figure, we can see that the range of e and Z$_{\rm max}$ are [0.898, 0.996] and [12~kpc, 200~kpc], respectively, and the
median values of $e$ and $Z_{\rm max}$ are 0.978 and 105~kpc, respectively. These results represent that the 246 HiVelSCs stars are on
highly eccentric orbits, and all of them have $e$ above 0.6 and $Z\rm _{max} > 3~kpc$, which represents that they are
kinematically consistent with the halo population \citep{2015MNRAS.447.2046H, 2013A&A...553A..19B, 2013MNRAS.436.3231K, 2013A&A...555A..12K}.
Table~\ref{tab:kinematic} lists $e$, $Z_{\rm max}$, $r_{\rm min}$, and $E-\Phi(\infty)$ in the last four columns, and Table~\ref{tab:general_catalog}
explains the four orbit parameters in detail. For unbound HiVelSCs and bound HiVelSCs with $Z_{\rm max} > 200$ kpc, Table~\ref{tab:kinematic} and~\ref{tab:general_catalog}
does not provide e and Z$_{\rm max}$.

\begin{figure*}
\begin{center}
\includegraphics[scale=0.4,angle=0]{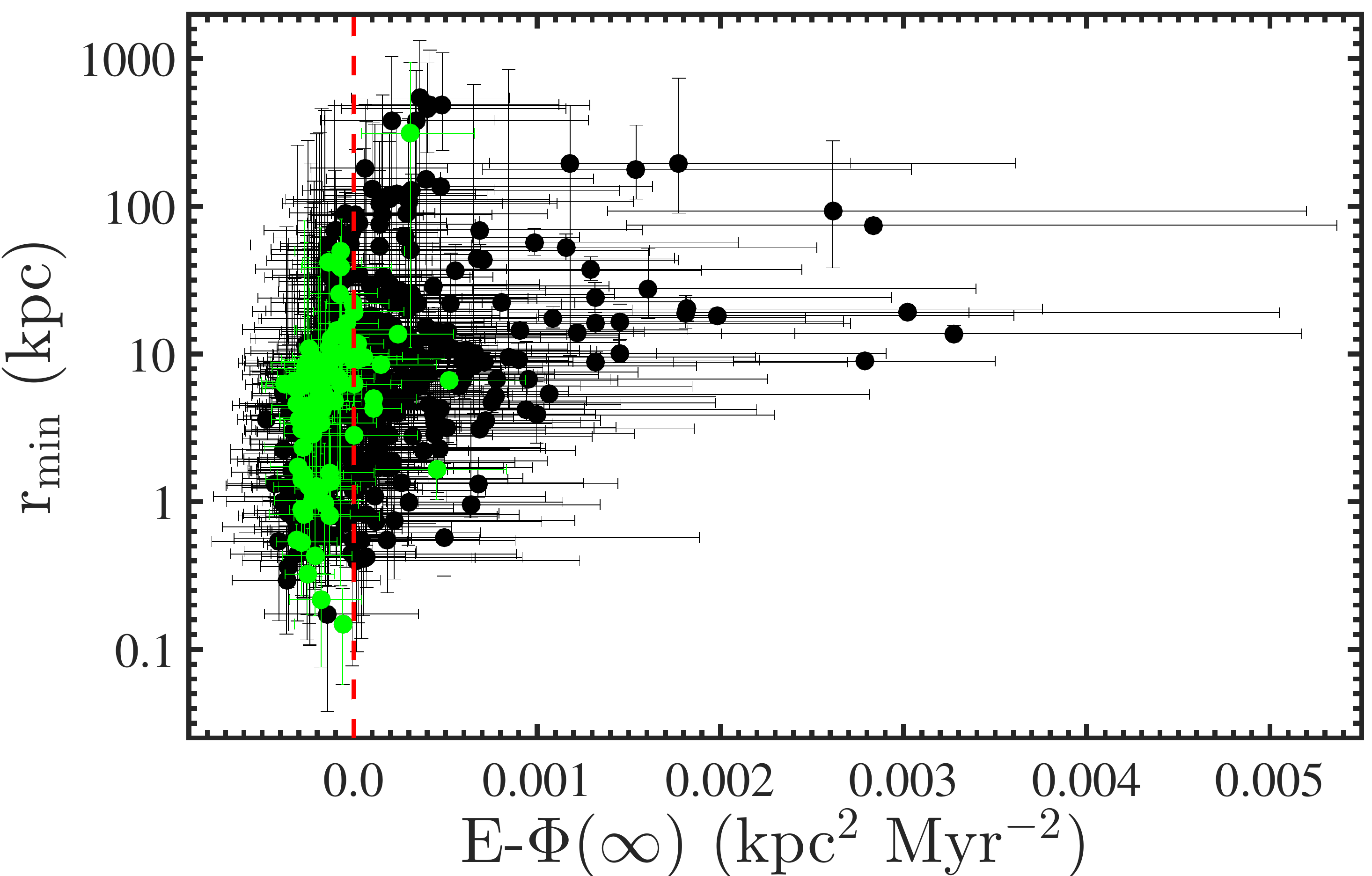}
\caption{Minimum crossing radius $r_{\rm min}$ as a function of energy $E-\Phi(\infty)$, and the y-axis is in log scale. The
dashed red line separates unbound ($E-\Phi(\infty) > 0$) from bound ($E-\Phi(\infty) < 0$) orbits.
The black and green solid dots are 591 HiVelSCs, and the green solid dots represent 92
conservative HiVelSCs introduced in Section~\ref{sec: other_discussion}. \label{fig:Rmin_E}}
\end{center}
\end{figure*}

In Figure~\ref{fig:Rmin_E}, we plot $r_{\rm min}$ as a function of the energy $E-\Phi(\infty)$, and the red dashed line
is the separation region between bound and unbound HiVelSCs. In this figure, there are 240 (41$\%$) HiVelSCs with remarkably
high values of the energy and traveling on unbound orbits ($E-\Phi(\infty) > 0$).

After tracing the orbit back, we derive the origins of 591 HiVelSCs from the positions of these
stars crossing the disk \citep{2019MNRAS.490..157M}. Using the minimum crossing radius $r_{min}$,
we obtain the probability that a star is ejected from the GC ($P_{\rm gc}$) which is defined as the fraction of orbits with $r_{\rm min} < 1$~kpc in 1000 MC orbits,
and the probability that a star is ejected from the Galaxy ($P_{\rm MW}$) that is defined as the fraction of orbits with $r_{\rm min} < 25$~kpc \citep{2015ApJ...801..105X, 2018ApJ...869L..31D, 2019ApJS..244....4D}. Using $P_{\rm gc}$, $P_{\rm MW}$, and the unbound probability of $P_{\rm ub}$ defined in Section~\ref{sec:candidates}, the HiVelSCs are divided into four types, and the classified criteria are shown in Table~\ref{tab:origin}. In this table, ``HVS'' represents the fastest hypervelocity stars in the Galaxy; ``HRS'' represents hyper-runaway stars; ``RS'' represents the runaway stars; and ``OUT'' represents fast halo stars. In our ``LAMOSTDR7-GaiaDR2-HiVelSC'' catalog, the column ``Origin$\_$Class'' records the result of this classification, and the last column of the Table~\ref{tab:unbound} lists the classification results for the 20 fastest HiVelSCs. Following the classification criteria in Table~\ref{tab:origin}, our HiVelSCs are divided into 91 ``HVS'' candidates, 107 ``HRS'' candidates, 221 ``RS'' candidates, and 172 ``OUT'' candidates. The ``HVS'' and ``OUT'' candidates are considered to be stripped from the GC and dwarf galaxies, respectively, and both ``HRS'' and ``RS'' stars were ejected from the Galactic disk.

The origins of our HiVelSCs from their tracing back orbits might suggest that a fraction (71$\%$) of them were ejected from the Galactic disk or bulge, and others were originated from dwarf galaxies. This appears to be in contradiction with that inferred from the metallicities of these HiVelSCs, i.e., only a small fraction (14$\%$) HiVelSCs were originated from the Galactic disk or bulge as explained in Section~\ref{sec:halo_disk}. The main reason for such a difference is that although the orbits of these stars can cross the Galactic disk or bulge, it does not necessarily mean that they must be ejected from disk or bulge. Besides, uncertainties of kinematic parameters and the selection of Galactic potential model will affect the orbits.

\begin{deluxetable*}{cccc}
\tablecaption{The classification criteria of HiVelSCs \label{tab:origin}}
\tablewidth{0pt}
\tablehead{
\colhead{Class} & \colhead{$P_{\rm gc}$} & \colhead{$P_{\rm MW}$} & \colhead{$P_{\rm ub}$} \\
}
\startdata
~~~~~~~~HVS candidates~~~~~~~~&~~~~~~~~$>$ 0.16~~~~~~~~&~~~~~~~~$-$~~~~~~~~&~~~~~~~~$-$~~~~~~~~   \\
~~~~~~~~HRS candidates~~~~~~~~&~~~~~~~~$<$ 0.16~~~~~~~~&~~~~~~~~$>$ 0.5~~~~~~~~&~~~~~~~~$>$ 0.5~~~~~~~~   \\
~~~~~~~~RS candidates~~~~~~~~~&~~~~~~~~$<$ 0.16~~~~~~~~&~~~~~~~~$>$ 0.5~~~~~~~~&~~~~~~~~$<$ 0.5~~~~~~~~  \\
~~~~~~~~OUT candidates~~~~~~~~&~~~~~~~~$<$ 0.16~~~~~~~~&~~~~~~~~$<$ 0.5~~~~~~~~&~~~~~~~~$-$~~~~~~~~   \\
\enddata
\end{deluxetable*}

\begin{figure*}
\begin{center}
\includegraphics[scale=0.45,angle=0]{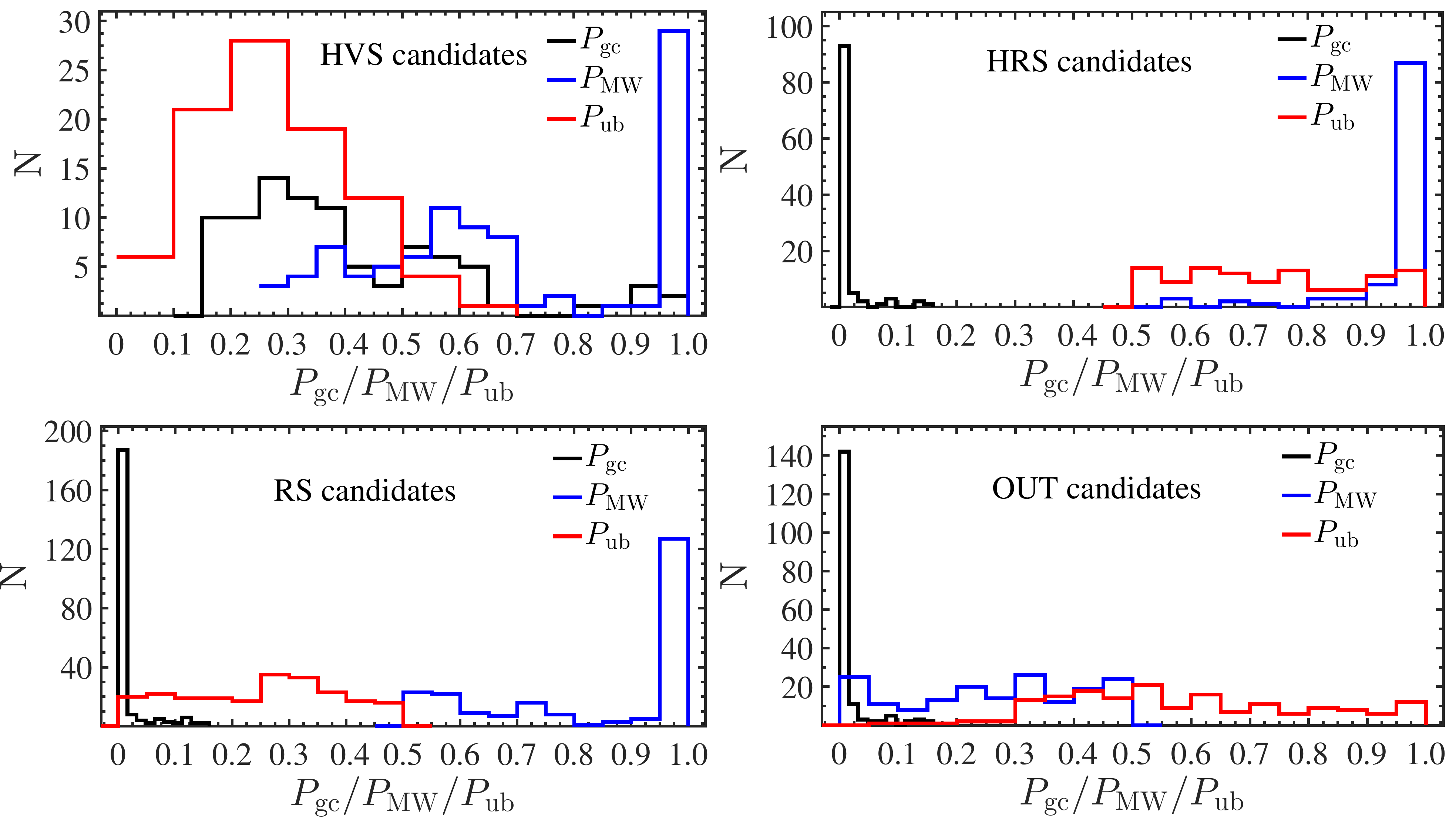}
\caption{The distribution of probability that a star is ejected from the GC ($P_{\rm gc}$), probability that a star is ejected from the Galaxy ($P_{\rm MW}$), and unbound
probability ($P_{\rm ub}$) for hypervelocity stars (``HVS''), hyper-runaway stars (``HRS''), runaway stars (``RS''), and fast halo stars (``OUT''), respectively. $P_{\rm gc}$ and
$P_{\rm MW}$ are defined in Section~\ref{sec:orbit}, and $P_{\rm ub}$ is defined in Section~\ref{sec:candidates}. \label{fig:origin_probability}}
\end{center}
\end{figure*}

In Figure~\ref{fig:origin_probability}, we plot the distribution of $P_{\rm gc}$, $P_{\rm MW}$, and $P_{\rm ub}$
for ``HVS'', ``HRS'', ``RS'', and ``OUT'' candidates, respectively. In the upper left panel, most ``HVS'' candidates are located in the region with
$P_{\rm gc} \leq 0.5$, $P_{\rm ub} \leq 0.5$ and $P_{\rm MW} > 0.5$, and it implies that our criterion might be too optimistic in the sense
that we might over-estimate the number of "HVS" stars. A fraction of "HVS" candidates will be proved to be other three types of stars
(``HRS'', ``RS'' or ``OUT'' stars). A similar conclusion can be drawn from other three panels of this figure, which indicates that the
classification criteria in Table~\ref{tab:origin} used in this work and other literature \citep{2018ApJ...869L..31D, 2019ApJS..244....4D, 2019MNRAS.490..157M}
might not be sufficient to determine the origin places of our HiVelSCs. In near future, more precise measurements of atmospheric parameters (in particular the metallicity, parallax, proper motions and improved Galactic potential models) will be available, a detailed orbit analysis to further investigate the origin places for our HiVelSCs will be carried
out as needed.

\section{Discussion} \label{sec:discussion&conclusion}
\subsection{The Zero-Point Correction of Parallax} \label{sec:parallax_zero_point_correct}

In this Section, we discuss the impact of a -0.029 mas global parallax zero-point mentioned in
\citet{2018A&A...616A...2L} for 591 HVCSs, derived from distant quasars. Being a negative offset, it
leads to lower inferred distance, and therefore smaller total velocity. We repeat the
calculations described in Section~\ref{sec:distance_velocity} but using the mean vector
$m = [\alpha, \delta, \varpi + \varpi_{\rm zp}, \mu_{\alpha^{*}}, \mu_{\delta}$]
to estimate Galactocentric distance and total velocity, where $\rm \varpi_{zp} = 0.029~mas$.

After such a parallax zero-point correction, both Galactocentric distances $r_{\rm GC}$ and total velocities$V_{\rm GC}$ are reduced for 591 HiVelSCs. The maximum, minimum, and mean value of $r_{\rm GC}$ reduction are
4.0~kpc, 0.2~pc, and 0.4~kpc, respectively, and those of $V_{\rm GC}$ reduction are 241~km s$^{-1}$, 0.15~km s$^{-1}$,
and 66~km s$^{-1}$ respectively. Unbound probabilities in the seven potential models
mentioned in Section~\ref{sec:candidates} are also re-estimated, and we find that the number of
unbound HiVelSCs in Watkins+2019 potential model with $P_{\rm ub} \geq 50\%$ decreased from 43 to 8, the number of bound HiVelSCs
with $P_{\rm ub} < 50\%$ in Xue+2008 potential model increase from 284 to 492, and whether other 99 HiVelSCs are unbound with $P_{\rm ub} \geq 50\%$
or not depend on the potential model used.

\subsection{Selecting HiVelSCs with More Conservative Criteria} \label{sec: other_discussion}

In Section~\ref{sec:sample}, we describe the selection criteria of 591 HiVelSCs, and here we focus on more stringent
selection criteria and show the corresponding results in Table~\ref{tab:strict_criteria}. The first column of each row in this table lists the selection criteria, and the second to sixth columns show the corresponding results of using these criteria, which are the total number ($\leq 591$) of the HiVelSCs,
the number of  HiVelSCs unbound in Watkins+2019 potential model with $P_{\rm ub} \geq 50 \%$,
the number of HiVelSCs unbound in the Xue+2008 potential model with $P_{\rm ub} \geq 50 \%$,
the number of HiVelSCs bound in the Xue+2008 potential model with $P_{\rm ub} < 50 \%$, and the maximum value of total Galactocentric
velocities for HiVelSCs selected by the criteria in the first column, respectively. As a comparison, the results using the selection
method in Section~\ref{sec:sample} are listed in the first row, and the results using other more conservative criteria are
listed from the second to fifth rows.

From the second row of Table~\ref{tab:strict_criteria}, we can see that 476 HiVelSCs have conservative high quality astrometric
parameters, i.e., astrometric$\_$flag = 1, and 34 of them are still unbound in Watkins+2019 potential model with $P_{\rm ub} \geq 50 \%$.
If we focus on HiVelSCs with more precise parallaxes, from the third row we can see that only 120 HiVelSCs satisfy this criteria, and only
two of them are unbound in Watkins+2019 potential model with $P_{\rm ub} \geq 50 \%$. If we focus on HiVelSCs with more precise total
velocities V$_{\rm GC}$, from the fourth row we can see that only 247 HiVelSCs are left, and three of them are unbound in Watkins+2019 potential model
with $P_{\rm ub} \geq 50 \%$.

Finally, if we use all criteria listed in the first column from the second to fourth rows, we can
see in the fifth row that only 92 HiVelSCs are left, and none of them are unbound in Watkins+2019 potential model with
$P_{\rm ub} \geq 50 \%$, but there are 26 HiVelSCs unbound in Xue+2008 potential model with $P_{\rm ub} \geq 50 \%$. In Figures~\ref{fig:gl_gb},
\ref{fig:Gmag}, \ref{fig:cmd}, \ref{fig:r_vgal}, \ref{fig:XYZ}, \ref{fig:vxvyvz}, \ref{fig:toomre}, \ref{fig:alpha_fe}, \ref{fig:kinematics_chemistry},
\ref{fig:zmax_eccentricity}, and \ref{fig:Rmin_E}, we also show the distribution of 92 conservative HiVelSCs.
Besides, in our ``LAMOSTDR7-GAIADR2-HiVelSC'' catalogue, the ``conservative$\_$result'' column is used to select the 92 conservative HiVelSCs by
using ``conservative$\_$result = 1''.

\begin{deluxetable*}{llllll}
\tablecaption{The results using more stringent selection criteria\label{tab:strict_criteria}}
\tablewidth{0pt}
\tablehead{
\colhead{Criteria $^{a}$} & \colhead{N1 $^{b}$} & N2$^{c}$ & N3$^{d}$ & N4$^{e}$ & \colhead{Maximum V$_{\rm GC}^{f}$} \\
\colhead{} & \colhead{} & \colhead{} & \colhead{} & \colhead{} & \colhead{(km s$^{-1}$)}
}
\startdata
only criteria mentioned in Sect. 2.3 used & 591 & 43 & 304 & 287 & 922 \\
astrometric$\_$flag = 1 & 476 & 34 & 245 & 231 & 888 \\
$f_{\rm \varpi} = \sigma_{\varpi} / \varpi \leq 0.1$ & 120 & 2 & 38 & 82 & 876 \\
$\rm \sigma_{V_{GC}} / V_{\rm GC} \leq$ 0.2 & 247 & 3 & 87 & 160 & 876 \\
above criteria are all used & 92 & 0 & 26 & 66 & 595
\enddata
$^{a}$ Selection criteria. The criteria from the second to fifth rows are adopted after using the criteria in the first row.
$^{b}$ Total number of HiVelSCs satisfied the selection criteria listed in the first column
$^{c}$ The number of HiVelSCs unbound in Watkins+2019 potential model with $P_{\rm ub} \geq 50\%$;
$^{d}$ The number of HiVelSCs unbound in Xue+2008 potential model with $P_{\rm ub} \geq 50\%$;
$^{e}$ The number of HiVelSCs bound in Xue+2008 potential model with $P_{\rm ub} < 50\%$;
$^{f}$ The maximum Galactocentric velocities for all HiVelSCs selected by the criteria in the first column;
\end{deluxetable*}

\section{Conclusion} \label{sec:conclusion}

In this paper, we cross-match LAMOST DR7 with Gaia DR2 to find more high velocity stars, and obtain a catalog
consisting of over 10 million spectra (LAMOST-Gaia). Galactocentric distances {$r_{\rm GC}$ and total velocities
$V_{\rm GC}$ are estimated for over 8.48 million ``low-f'' entries in the LAMOST-Gaia catalog, which have
positive Gaia parallax and low fractional parallax errors ($f_{\rm \varpi} = \sigma_{\rm \varpi} / \varpi \leq 0.2$).
We define our sample of high velocity stars as those, who can escape from the Galaxy in at least one of the
seven Galactic potential models adopted in this work, or have total velocities in the Galactic rest frame larger
than 450 km s$^{-1}$, and result in a total of 591 high velocity star candidates (HiVelSCs) found from the LAMOST-Gaia catalog.
Their total velocities $V_{\rm GC}$ are then recalculated using the radial velocities determined by the
LAMOST Stellar Parameter Pipeline (LASP), because their $RV$ uncertainties given by the LASP not only consider the calculation
error caused by LASP algorithm but also take into account the error introduced from the process of spectra observation
and data processing. After the recalculation, the total velocities for all HiVelSCs are still $> 445$~km s$^{-1}$, and our main
conclusions are:

(i)  Among 591 HVCs, the majority of them are giant stars, and 14 of them have been already collected
in the open fast stars catalog (OFSC) \citep{2018MNRAS.479.2789B}. In this paper, we construct a catalogue of ``LAMOSTDR7-GAIADR2-HiVelSC''
to collect parameters for 591 HiVelSCs, including 93 columns such as LAMOST radial velocity, atmospheric parameters, and five Gaia
astrometry parameters, and 476 HiVelSCs have conservative high quality astrometric parameters of Gaia DR2,
which satisfy the criteria introduced in Section~\ref{sec:candidates}.

(ii) Using seven Galactic potential models, 43 HiVelSCs can escape from the Galaxy
with unbound probabilities of $P_{\rm ub} > 50\%$ in the potential model of \citet{2019ApJ...873..118W} (Watkins+2019; resulting in the largest escaping velocities),
304 HiVelSCs are unbound in the potential model of \citet{2008ApJ...684.1143X} (Xue+2008; resulting in the smallest escaping velocities) with $P_{\rm ub} > 50\%$, and other 287 HiVelSCs are bound in Xue+2008 potential model with P$_{\rm ub} \leq 50\%$. As mentioned in \citet{2018A&A...616A...2L} and \citet{2018A&A...616A..17A}, there is a
negative zero-point for Gaia parallax, and we attempt to correct the zero-point of -0.029 for 591 HiVelSCs, which leads to a decrease in the number
of HiVelSCs unbound in Watkins+2019 potential model with P$_{\rm ub} > 50\%$, from 43 to 8. At the end of this work, we also discuss the effect if more conservative criteria are adopted to select HiVelSCs. The conclusion is that there are in total 92 HiVelSCs satisfying all of the three more conservative selection criteria, and they are referred to as 92 conservative HiVelSCs in the main text. Among them, there is no HiVelSCs unbound in Watkins+2019 potential model with P$_{\rm ub} > 50\%$ when all criteria listed in Table~\ref{tab:strict_criteria} are used, but there are 26 HiVelSCs have P$_{\rm ub} > 50\%$ of being unbound in Xue+2019 potential model, and the maximum value of Galactocentric velocities $V_{\rm GC}$ for 92 conservative HiVelSCs is 595 km s$^{-1}$.

(iii) Using the Toomre diagram, probabilities for the thick-disk-to-halo (TD/H) and the distribution on the (eccentricity, $z_{\rm max}$) plane,
we find that all 591 HiVelSCs are kinematically associated to the halo population, which confirms the assumptions, that high velocity stars in the solar vicinity mostly belong to the halo, adopted in previous works \citep{1988A&AS...73..225S, 2003MNRAS.341..199R, 2006A&A...445..939S}. The 591 HiVelSCs have a
mean $\alpha$ abundance of $\alpha = +0.22$ dex, and their metallicities [Fe/H] peak at near [Fe/H]~$\sim$~$-$1.2 and have a wide range
from near [Fe/H]~$\sim$~$-$3.5 to [Fe/H]~$\sim$~+0.5, which also indicates that on the whole they are metal-poor slightly
$\alpha$-enhanced inner halo stars.

Among the 591 HiVelSCs, about 14$\%$ of them are metal-rich halo stars with [Fe/H]$> -$1, and they
may be the in situ halo stars \citep{2010A&A...511L..10N, 2019ApJ...887..237C}. These in situ halo stars are
considered to be formed in the initial collapse of the Milky Way or formed in the disk or bulge and be kinematically heated subsequently
\citep{2019ApJ...887..237C, 2017ApJ...845..101B, 2018ApJ...863...87D}, and the low ratio of them indicates that the bulk of the stellar 
halo formed as a consequence of the accretion and tidal disruption processes as mentioned in \citet{2019ApJ...887..237C}.

(iv) Considering both their kinematics and chemistries, we find that most of our HiVelSCs have retrograde velocities of V$_{\rm y}$, and
retrograde HiVelSCs have a slightly lower mean value of [Fe/H], which is consistent with the results in the literature \citep{2017ApJ...841...59Z, 2018ApJ...862..163L, 2019ApJ...887..237C}, meanwhile metal-poor ([Fe/H] $<$ -1) HiVelSCs tend to have faster mean retrograde velocities $V_{\rm y}$.

(v) There are 304 HiVelSCs unbound in Xue+2008 potential model with $P_{\rm ub} > 50\%$, and most of them move on
retrograde ($V_{\rm y} < 0$) orbit, which means, as it has been suggested in the literature, that a population of unbound stars may have
an extragalactic provenance \citep{2019A&A...627A.104D, 2019MNRAS.490..157M, 2016A&A...588A..41C}. In order to further investigate their origin, we track the
orbits of 591 HiVelSCs back in time, and they are divided into four types, i.e., ``HVS'', ``HRS'', ``RS'' and ``OUT'', according to their unbound
probabilities ($P_{\rm ub}$), and the probabilities of being ejected from the GC ($P_{\rm gc}$) and Galactic disk ($P_{\rm MW}$).
We find that $\sim$15$\%$ of stars are from the GC, $\sim$55$\%$ of stars are from the Galactic disk, and $\sim$30$\%$ of stars have extragalactic origins,
but we caution that due to the uncertainty in the phase space measurement, further investigation is needed to confirm their origins.

\appendix
\renewcommand\thetable{\Alph{section}\arabic{table}}

\section{Tables} \label{sec:appendix}

Here, we list four tables. Table~\ref{tab:vesc} and \ref{tab:unbound} list escape velocities $V_{\rm esc}$ and unbound probabilities $P_{\rm ub}$ (defined in Section~\ref{sec:candidates} ) for the 20 fastest HiVelSCs in the seven potential models \citep{1990ApJ...348..485P, 2005ApJ...634..344G, 2008ApJ...684.1143X, 2010ApJ...712..260K, 2014ApJ...793..122K, 2015ApJS..216...29B, 2019ApJ...873..118W} mentioned in Section~\ref{sec:candidates}, and Table~\ref{tab:atmospheric} lists their atmospheric parameters (effective temperature $T_{\rm eff}$, surface gravity log~$g$, and metallicity [Fe/H]) and $\alpha$ element abundances ([$\alpha$/Fe]). As mentioned in Section \ref{sec:candidates}, we construct the catalogue of ``LAMOSTDR7-GAIADR2-HiVelSC'' including 93 columns to list various parameters for 591 HiVelSCs, and Table~\ref{tab:general_catalog} explains each column of this table in detail.

\setcounter{table}{0}
\renewcommand{\thetable}{A\arabic{table}}

\begin{deluxetable*}{cccccccccc}
\caption{Escape velocities of the 20 fastest HiVelSCs in the seven Galactic potential models\label{tab:vesc}}
\tablewidth{0pt}
\tablehead{
\colhead{ID} & \colhead{$r_{\rm GC}$$^{a}$} & \colhead{$V_{\rm GC}$$^{a}$} & \colhead{$V_{\rm esc}$(W)$^{b}$} & \colhead{$V_{\rm esc}$(G)$^{b}$} & \colhead{$V_{\rm esc}$(Ke)$^{b}$} & \colhead{$V_{\rm esc}$(Ko)$^{b}$} & \colhead{$V_{\rm esc}$(P)$^{b}$} & \colhead{$V_{\rm esc}$(M)$^{b}$} & \colhead{$V_{\rm esc}$(X)$^{b}$}  \\
\colhead{} & \colhead{(kpc)} & \colhead{(km s$^{-1}$)} & \colhead{(km s$^{-1}$)} & \colhead{(km s$^{-1}$)} & \colhead{(km s$^{-1}$)} & \colhead{(km s$^{-1}$)} & \colhead{(km s$^{-1}$)} & \colhead{(km s$^{-1}$)} & \colhead{(km s$^{-1}$)}
}
\startdata
Hivel1 & 10.1$_{-0.6}^{+0.9}$ & 922$_{-136}^{+168}$ & 598 & 587 & 573 & 543 & 512 & 491 & 474  \\
Hivel2 & 10.6$_{-0.4}^{+0.6}$ & 888$_{-142}^{+197}$ & 595 & 585 & 570 & 542 & 511 & 486 & 469  \\
Hivel3 & 8.9$_{-0.0}^{+0.1}$ & 876$_{-73}^{+81}$ & 615 & 601 & 590 & 567 & 533 & 503 & 483  \\
Hivel4 & 10.9$_{-0.5}^{+0.8}$ & 874$_{-168}^{+247}$ & 593 & 584 & 568 & 540 & 509 & 484 & 468  \\
Hivel5 & 11.5$_{-0.9}^{+1.6}$ & 862$_{-147}^{+242}$ & 585 & 579 & 561 & 530 & 501 & 479 & 464  \\
Hivel6 & 6.3$_{-0.4}^{+0.8}$ & 791$_{-142}^{+199}$ & 634 & 617 & 610 & 576 & 544 & 535 & 508  \\
Hivel7 & 8.8$_{-0.2}^{+0.4}$ & 789$_{-126}^{+194}$ & 611 & 598 & 586 & 557 & 525 & 504 & 484  \\
Hivel8 & 7.9$_{-0.0}^{+0.0}$ & 778$_{-69}^{+81}$ & 624 & 608 & 600 & 573 & 540 & 514 & 492  \\
Hivel9 & 10.6$_{-1.0}^{+1.7}$ & 741$_{-156}^{+236}$ & 592 & 583 & 568 & 537 & 507 & 486 & 469  \\
Hivel10 & 7.7$_{-0.1}^{+0.3}$ & 737$_{-114}^{+174}$ & 623 & 606 & 599 & 569 & 537 & 516 & 493  \\
Hivel11 & 15.2$_{-1.2}^{+2.0}$ & 727$_{-123}^{+186}$ & 561 & 558 & 535 & 507 & 480 & 453 & 442  \\
Hivel12 & 12.5$_{-1.0}^{+1.8}$ & 726$_{-138}^{+230}$ & 578 & 573 & 553 & 524 & 495 & 471 & 457  \\
Hivel13 & 6.4$_{-0.1}^{+0.3}$ & 718$_{-87}^{+121}$ & 636 & 616 & 612 & 580 & 548 & 533 & 506  \\
Hivel14 & 11.6$_{-1.2}^{+2.0}$ & 713$_{-108}^{+160}$ & 588 & 575 & 563 & 537 & 506 & 478 & 462  \\
Hivel15 & 8.1$_{-0.1}^{+0.3}$ & 705$_{-121}^{+191}$ & 619 & 603 & 594 & 566 & 534 & 511 & 489  \\
Hivel16 & 9.0$_{-0.3}^{+0.5}$ & 698$_{-104}^{+132}$ & 607 & 596 & 583 & 552 & 521 & 502 & 482  \\
Hivel17 & 6.6$_{-0.1}^{+0.2}$ & 697$_{-137}^{+165}$ & 637 & 617 & 612 & 583 & 550 & 531 & 504  \\
Hivel18 & 9.3$_{-0.6}^{+1.1}$ & 697$_{-110}^{+170}$ & 606 & 592 & 581 & 552 & 520 & 499 & 479  \\
Hivel19 & 9.3$_{-0.3}^{+0.6}$ & 696$_{-131}^{+198}$ & 606 & 594 & 581 & 552 & 521 & 499 & 479  \\
Hivel20 & 10.5$_{-0.4}^{+0.6}$ & 691$_{-118}^{+149}$ & 596 & 586 & 571 & 542 & 512 & 488 & 471  \\
\enddata
\tablecomments{
The 62th to 68th columns of Table~\ref{tab:general_catalog} are listed here for the 20 fastest HiVelSCs,
which mainly includes the escape velocities estimated by the seven Galactic potential models
\citep{1990ApJ...348..485P, 2005ApJ...634..344G, 2008ApJ...684.1143X, 2010ApJ...712..260K, 2014ApJ...793..122K, 2015ApJS..216...29B, 2019ApJ...873..118W}.
The median value, lower uncertainty, and upper uncertainty of Galactocentric distance ($r_{\rm GC}$)
are shown in the same column (the second column) in this table but separated into three columns, respectively, in Table~\ref{tab:general_catalog}
(from 35th to 37th columns). The median value, lower uncertainty, and upper uncertainty of Galactocentric total velocity ($V_{\rm GC}$)
are shown in the same column (the third column) in this table but separated into three columns, respectively, in Table~\ref{tab:general_catalog}
(from 47th to 49th columns). $^{a}$ Galactocentric distance and total velocity; $^{b}$ Escape velocities in the seven Galactic potential models;
}
\end{deluxetable*}

\begin{deluxetable*}{ccccccccccc}
\caption{Unbound probabilities of the 20 fastest HiVelSCs in the seven potential models\label{tab:unbound}}
\tablewidth{0pt}
\tablehead{
\colhead{ID} & \colhead{$P_{\rm gc}$$^{a}$} & \colhead{$P_{\rm MW}$$^{b}$} & \colhead{$P_{\rm ub}$(W)$^{c}$} & \colhead{$P_{\rm ub}$(G)$^{c}$} &
\colhead{$P_{\rm ub}$(Ke)$^{c}$} & \colhead{$P_{\rm ub}$(Ko)$^{c}$} & \colhead{$P_{\rm ub}$(P)$^{c}$} & \colhead{$P_{\rm ub}$(M)$^{c}$} &
\colhead{$P_{\rm ub}$(X)$^{c}$} & \colhead{Origin$\_$Class$^{d}$}   \\
\colhead{} & \colhead{} & \colhead{} & \colhead{} & \colhead{} & \colhead{} & \colhead{} & \colhead{} & \colhead{} & \colhead{} & \colhead{}
}
\startdata
Hivel1 & 0.000 & 1.000 & 0.996 & 0.998 & 1.000 & 1.000 & 1.000 & 1.000 & 1.000 & HRS  \\
Hivel2 & 0.000 & 0.993 & 0.995 & 0.999 & 1.000 & 1.000 & 1.000 & 1.000 & 1.000 & HRS  \\
Hivel3 & 0.000 & 1.000 & 1.000 & 1.000 & 1.000 & 1.000 & 1.000 & 1.000 & 1.000 & HRS  \\
Hivel4 & 0.000 & 0.039 & 0.969 & 0.973 & 0.978 & 0.991 & 0.996 & 0.999 & 0.999 & OUT  \\
Hivel5 & 0.000 & 0.000 & 0.975 & 0.981 & 0.984 & 0.990 & 0.997 & 0.999 & 1.000 & OUT  \\
Hivel6 & 0.000 & 0.002 & 0.861 & 0.898 & 0.904 & 0.951 & 0.981 & 0.989 & 0.997 & OUT  \\
Hivel7 & 0.000 & 0.994 & 0.928 & 0.947 & 0.953 & 0.981 & 0.995 & 0.998 & 1.000 & HRS  \\
Hivel8 & 0.000 & 0.000 & 0.992 & 0.997 & 0.999 & 1.000 & 1.000 & 1.000 & 1.000 & OUT  \\
Hivel9 & 0.000 & 0.852 & 0.804 & 0.831 & 0.852 & 0.912 & 0.935 & 0.951 & 0.969 & HRS  \\
Hivel10 & 0.001 & 0.003 & 0.837 & 0.874 & 0.894 & 0.939 & 0.972 & 0.989 & 0.999 & OUT  \\
Hivel11 & 0.000 & 0.848 & 0.900 & 0.913 & 0.940 & 0.961 & 0.983 & 0.994 & 0.996 & HRS  \\
Hivel12 & 0.000 & 0.415 & 0.843 & 0.854 & 0.886 & 0.929 & 0.965 & 0.980 & 0.988 & OUT  \\
Hivel13 & 0.000 & 0.000 & 0.821 & 0.868 & 0.876 & 0.942 & 0.979 & 0.987 & 0.997 & OUT  \\
Hivel14 & 0.000 & 0.997 & 0.863 & 0.885 & 0.910 & 0.942 & 0.982 & 0.994 & 0.996 & HRS  \\
Hivel15 & 0.000 & 0.995 & 0.759 & 0.790 & 0.809 & 0.871 & 0.931 & 0.959 & 0.978 & HRS  \\
Hivel16 & 0.000 & 0.009 & 0.800 & 0.830 & 0.855 & 0.923 & 0.965 & 0.984 & 0.993 & OUT  \\
Hivel17 & 0.000 & 1.000 & 0.659 & 0.713 & 0.732 & 0.797 & 0.864 & 0.915 & 0.953 & HRS  \\
Hivel18 & 0.000 & 1.000 & 0.782 & 0.816 & 0.839 & 0.908 & 0.961 & 0.979 & 0.990 & HRS  \\
Hivel19 & 0.000 & 0.022 & 0.736 & 0.775 & 0.798 & 0.857 & 0.909 & 0.938 & 0.962 & OUT  \\
Hivel20 & 0.000 & 0.999 & 0.779 & 0.809 & 0.837 & 0.904 & 0.946 & 0.975 & 0.983 & HRS  \\
\enddata
\tablecomments{
The 69th to 78th columns of Table~\ref{tab:general_catalog} are listed here for the 20 fastest HiVelSCs,
which include the unbound probabilities in the seven potential models
\citep{1990ApJ...348..485P, 2005ApJ...634..344G, 2008ApJ...684.1143X, 2010ApJ...712..260K, 2014ApJ...793..122K, 2015ApJS..216...29B, 2019ApJ...873..118W},
and the classification results of their origin.
$^{a}$ The probability that a star is derived from the GC;
$^{b}$ The probability that a star origins from the Galactic disk;
$^{c}$ The unbound probabilities in the seven Galactic potential models;
$^{d}$ ;
}
\end{deluxetable*}

\begin{longrotatetable}
\begin{deluxetable*}{cccccccc}
\tablecaption{Atmospheric parameters for the 20 fastest HiVelSCs\label{tab:atmospheric}}
\tablewidth{0pt}
\tablehead{
\colhead{ID} & \colhead{$T_{\rm eff}$\_LASP$^{a}$} & \colhead{log~$g$\_LASP$^{a}$} & \colhead{[Fe/H]$\_$LASP$^{a}$} & \colhead{$T_{\rm eff}$\_DD-Payne$^{b}$} & \colhead{log~$g$\_DD-Payne$^{b}$} & \colhead{[Fe/H]$\_$DD-Payne$^{b}$} &\colhead{[$\alpha$/Fe]$\_$DD-Payne$^{b}$} \\
\colhead{} & \colhead{(K)} & \colhead{(dex)} & \colhead{(dex)} & \colhead{(K)} & \colhead{(dex)} & \colhead{(dex)} & \colhead{(dex)}
}
\startdata
Hivel1 & 4803$\pm$19 & 1.89$\pm$0.03 & -1.40$\pm$0.02 & 4919$\pm$38 & 37.62$\pm$2.22 & -1.25$\pm$0.06 & 0.28$\pm$0.02   \\
Hivel2 & 5384$\pm$84 & 2.82$\pm$0.14 & -1.54$\pm$0.08 & 5813$\pm$47 & 46.96$\pm$3.66 & -1.24$\pm$0.07 & 0.24$\pm$0.04   \\
Hivel3 & 4615$\pm$228 & 4.86$\pm$0.36 & -1.48$\pm$0.21 & 4801$\pm$51 & 50.95$\pm$4.93 & -1.09$\pm$0.09 & 0.17$\pm$0.04   \\
Hivel4 & 6553$\pm$12 & 4.10$\pm$0.02 & -2.30$\pm$0.02 & 7190$\pm$49 & 48.50$\pm$4.02 & -1.87$\pm$0.10 & 0.16$\pm$0.03   \\
Hivel5 & 4835$\pm$121 & 1.54$\pm$0.19 & -2.20$\pm$0.11 & 5366$\pm$81 & 80.92$\pm$2.37 & -1.84$\pm$0.13 & 0.22$\pm$0.10   \\
Hivel6 & 5415$\pm$32 & 2.78$\pm$0.05 & -1.10$\pm$0.03 & 5443$\pm$29 & 29.04$\pm$2.53 & -1.12$\pm$0.05 & 0.28$\pm$0.02   \\
Hivel7 & 6238$\pm$214 & 3.97$\pm$0.34 & -1.60$\pm$0.20 & 6175$\pm$90 & 89.67$\pm$3.35 & -1.88$\pm$0.18 & 0.30$\pm$0.07   \\
Hivel8 & 5401$\pm$29 & 4.16$\pm$0.05 & -0.46$\pm$0.03 & 5362$\pm$24 & 24.47$\pm$4.08 & -0.41$\pm$0.04 & 0.09$\pm$0.01   \\
Hivel9 & 4243$\pm$42 & 0.75$\pm$0.07 & -1.29$\pm$0.04 & 4290$\pm$21 & 21.40$\pm$1.17 & -1.22$\pm$0.04 & 0.10$\pm$0.02   \\
Hivel10 & 6151$\pm$260 & 4.06$\pm$0.41 & -0.79$\pm$0.24 & $-$ & $-$ & $-$ & -0.29$\pm$0.02   \\
Hivel11 & 4650$\pm$48 & 1.53$\pm$0.08 & -1.43$\pm$0.05 & 4798$\pm$50 & 49.53$\pm$1.92 & -1.25$\pm$0.07 & 0.27$\pm$0.03   \\
Hivel12 & 4928$\pm$26 & 2.13$\pm$0.04 & -1.34$\pm$0.02 & 5058$\pm$35 & 34.58$\pm$2.36 & -1.20$\pm$0.05 & 0.26$\pm$0.02  \\
Hivel13 & 5010$\pm$59 & 2.31$\pm$0.10 & -1.38$\pm$0.06 & 5054$\pm$57 & 57.30$\pm$2.32 & -1.28$\pm$0.08 & 0.26$\pm$0.03   \\
Hivel14 & 5013$\pm$39 & 2.26$\pm$0.06 & -1.97$\pm$0.04 & 5111$\pm$40 & 40.42$\pm$3.06 & -1.94$\pm$0.06 & 0.01$\pm$0.02  \\
Hivel15 & 6159$\pm$108 & 4.17$\pm$0.18 & -0.97$\pm$0.10 & 6118$\pm$36 & 36.28$\pm$4.14 & -1.07$\pm$0.07 & 0.18$\pm$0.04   \\
Hivel16 & 6336$\pm$29 & 4.14$\pm$0.05 & -1.21$\pm$0.03 & 6151$\pm$27 & 27.28$\pm$3.18 & -1.31$\pm$0.05 & 0.16$\pm$0.02   \\
Hivel17 & 6181$\pm$129 & 4.05$\pm$0.21 & -1.51$\pm$0.12 & 6220$\pm$55 & 54.78$\pm$3.95 & -1.55$\pm$0.10 & 0.18$\pm$0.06  \\
Hivel18 & 4660$\pm$50 & 1.40$\pm$0.08 & -2.04$\pm$0.05 & 5009$\pm$47 & 46.74$\pm$2.19 & -1.70$\pm$0.06 & 0.24$\pm$0.03  \\
Hivel19 & 4930$\pm$49 & 2.36$\pm$0.08 & -1.22$\pm$0.05 & 5024$\pm$50 & 49.61$\pm$2.52 & -1.16$\pm$0.07 & 0.25$\pm$0.03  \\
Hivel20 & 5271$\pm$95 & 2.76$\pm$0.16 & -2.21$\pm$0.09 & 5435$\pm$54 & 53.93$\pm$3.08 & -2.00$\pm$0.10 & 0.27$\pm$0.06  \\
\enddata
\tablecomments{
The 79th to 92nd columns of Table~\ref{tab:general_catalog} are listed here for the 20 fastest HiVelSCs, which are
atmospheric parameters and $\alpha$ abundance estimated by the the LASP and the method of
data-driven Payne (DD-Payne), respectively. The measurements and uncertainties are shown in the same column in this table but
separated into two columns in Table~\ref{tab:general_catalog}.
$^{a}$ Atomespheric parameters estimated by the LASP;
$^{b}$ Atomespheric parameters calculated by the method of data-driven Payne (DD-Payne) in \citet{2019ApJS..245...34X};
}
\end{deluxetable*}
\end{longrotatetable}

\begin{longtable}{cccc}

    \caption{Description of our HiVelSCs catalog. Parameters measured by this work such as distances and velocities correspond to the median of the distribution, and lower and upper uncertainties are derived respectively from the 16th and 84th percentiles of the distribution. Entries labelled$^{1}$ are from the LAMOST catalog, entries labelled$^{2}$ are taken from the Gaia DR2 catalogue \citep{2018A&A...616A...1G}, and entries labelled$^{3}$ are derived in this paper.}
    \label{tab:general_catalog} \\
    \hline
    Column & Name & Units & Description \\
    \hline
    \endfirsthead

    \multicolumn{4}{l}{Table A4$~$$-$$~$continued} \\
    \hline
    Column & Name & Units & Description \\
    \hline
    \endhead

    \hline
    \multicolumn{4}{c}{Continued on the next page}
    \endfoot

    \hline
    \endlastfoot

    1 & ID & $-$ & High Velocity star identifier $^{3}$ \\
    2 & specid & $-$ & LAMOST DR7 identifier $^{1}$ \\
    3 & Gaia\_designation & $-$ & Gaia DR2 identifier $^{2}$ \\
    4 & R.A. & deg & Right ascension $^{2}$ \\
    5 & e\_R.A. & deg & Right ascension error $^{2}$ \\
    6 & decl. & deg & Declination $^{2}$ \\
    7 & e\_decl. & deg & Declination error $^{2}$ \\
    8 & S/N\_r & $-$ & r-band spectral signal to noise ratio $^{1}$ \\
    9 & Class & $-$ & spectral type $^{1}$ \\
    10 & RV\_LASP & km~s$^{-1}$ & Radial velocity measured by the LASP $^{1}$ \\
    11 & e\_RV\_LASP & km~s$^{-1}$ & Radial velocity error given by the LASP $^{1}$ \\
    12 & pmra & mas~yr$^{-1}$ & Proper motion in right ascension $^{2}$ \\
    13 & e\_pmra & mas~yr$^{-1}$  & Standard uncertainty in pmra $^{2}$ \\
    14 & pmdec & mas~yr$^{-1}$ & Proper motion in declination $^{2}$ \\
    15 & e\_pmdec & mas~yr$^{-1}$ & Standard uncertainty in pmdec$^{2}$ \\
    16 & parallax & mas & Parallax $^{2}$ \\
    17 & e\_parallax & mas & Standard uncertainty in parallax $^{2}$ \\
    18 & G & mag  & G band mean magnitude $^{2}$ \\
    19 & G\_BP & mag  & G$\_$BP band mean magnitude $^{2}$ \\
    20 & G\_RP  & mag  & G$\_$RP band mean magnitude $^{2}$ \\
    21 & astrometric\_flag & $-$ & A flag to mark whether a star has reliable astrometric parameters $^{3}$ \\
    22 & RV\_LASP\_Calibrate & km~s$^{-1}$ & LASP radial velocity corrected the zero-point $^{1}$ \\
    23 & r & kpc & Heliocentric distance $^{3}$ \\
    24 & el\_r & kpc & Lower uncertainty on r $^{3}$ \\
    25 & eu\_r & kpc & Upper uncertainty on r $^{3}$ \\
    26 & x & kpc & Cartesian Galactocentric x coordinate $^{3}$ \\
    27 & el\_x & kpc & Lower uncertainty on x $^{3}$ \\
    28 & eu\_x & kpc & Upper uncertainty on x $^{3}$ \\
    29 & y & kpc & Cartesian Galactocentric y coordinate $^{3}$ \\
    30 & el\_y & kpc & Lower uncertainty on y $^{3}$ \\
    31 & eu\_y & kpc & Upper uncertainty on y $^{3}$ \\
    32 & z & kpc & Cartesian Galactocentric z coordinate$^{3}$ \\
    33 & el\_z & kpc & Lower uncertainty on z $^{3}$ \\
    34 & eu\_z & kpc & Upper uncertainty on z $^{3}$ \\
    35 & r\_GC & kpc & Galactocentric distance $^{3}$ \\
    36 & el\_r\_GC & kpc & Lower uncertainty on r$\_$GC $^{3}$ \\
    37 & eu\_r\_GC & kpc & Upper uncertainty on r$\_$GC $^{3}$ \\
    38 & Vx & km~s$^{-1}$ & Cartesian Galactocentric x velocity $^{3}$ \\
    39 & el\_Vx & km~s$^{-1}$ & Lower uncertainty on Vx $^{3}$ \\
    40 & eu\_Vx & km~s$^{-1}$ & Upper uncertainty on Vx $^{3}$ \\
    41 & Vy & km~s$^{-1}$ & Cartesian Galactocentric y velocity $^{3}$ \\
    42 & el\_Vy & km~s$^{-1}$ & Lower uncertainty on Vy $^{3}$ \\
    43 & eu\_Vy & km~s$^{-1}$ & Upper uncertainty on Vy $^{3}$ \\
    44 & Vz & km~s$^{-1}$ & Cartesian Galactocentric z velocity $^{3}$ \\
    45 & el\_Vz & km~s$^{-1}$ & Lower uncertainty on Vz $^{3}$ \\
    46 & eu\_Vz & km~s$^{-1}$ & Upper uncertainty on Vz $^{3}$ \\
    47 & V\_GC & km~s$^{-1}$ & Cartesian Galactocentric total velocity $^{3}$ \\
    48 & el\_V\_GC & km~s$^{-1}$ & Lower uncertainty on V$\_$GC $^{3}$ \\
    49 & eu\_V\_GC & km~s$^{-1}$ & Upper uncertainty on V$\_$GC $^{3}$ \\
    50 & e & $-$ & Orbit eccentricirty $^{3}$ \\
    51 & el\_e & $-$ & Lower uncertainty on e $^{3}$ \\
    52 & eu\_e & $-$ & Upper uncertainty on e $^{3}$ \\
    53 & Zmax & kpc & Orbit maximum height above the Galactic disk $^{3}$ \\
    54 & el\_Zmax & kpc & Lower uncertainty on Zmax $^{3}$ \\
    55 & eu\_Zmax & kpc & Upper uncertainty on Zmax $^{3}$ \\
    56 & r\_min & kpc &  Minimum crossing radius $^{3}$ \\
    57 & el\_r\_min & kpc & Lower uncertainty on r$\_$min $^{3}$ \\
    58 & eu\_r\_min & kpc & Upper uncertainty on r$\_$min $^{3}$ \\
    59 & E\_Phi\_Infinity & kpc$^{2}$~Myr$^{-2}$ &  The difference between orbit energy E and potential energy at infinity $^{3}$ \\
    60 & el\_E\_Phi\_Infinity & kpc$^{2}$~Myr$^{-2}$ & Lower uncertainty on E$\_$Phi$\_$Infinity $^{3}$ \\
    61 & eu\_E\_Phi\_Infinity & kpc$^{2}$~Myr$^{-2}$ & Upper uncertainty on E$\_$Phi$\_$Infinity $^{3}$\\
    62 & Vesc(W) & km~s$^{-1}$ & Escape velocity estimated by the potential model in \citet{2019ApJ...873..118W} $^{3}$\\
    63 & Vesc(G) & km~s$^{-1}$ & Escape velocity estimated by the potential model in \citet{2005ApJ...634..344G} $^{3}$ \\
    64 & Vesc(Ke) & km~s$^{-1}$ & Escape velocity estimated by the potential model in \citet{2014ApJ...793..122K} $^{3}$ \\
    65 & Vesc(Ko) & km~s$^{-1}$ & Escape velocity estimated by the potential model in \citet{2010ApJ...712..260K} $^{3}$ \\
    66 & Vesc(P) & km~s$^{-1}$ & Escape velocity estimated by the potential model in \citet{1990ApJ...348..485P} $^{3}$ \\
    67 & Vesc(M) & km~s$^{-1}$ & Escape velocity estimated by the potential model in \citet{2015ApJS..216...29B} $^{3}$ \\
    68 & Vesc(X) & km~s$^{-1}$ & Escape velocity estimated by the potential model in \citet{2008ApJ...684.1143X}  $^{3}$\\
    69 & Pgc & $-$ & Probability of ejecting from the GC $^{3}$ \\
    70 & P$_{\rm MW}$ & $-$ & Probability of ejecting from the Galactic disk $^{3}$ \\
    71 & Pub(W) & $-$ & Unbound probability in the potential model in \citet{2019ApJ...873..118W} $^{3}$ \\
    72 & Pub(G) & $-$ & Unbound probability in the potential model in \citet{2005ApJ...634..344G} $^{3}$ \\
    73 & Pub(Ke) & $-$ & Unbound probability in the potential model in \citet{2014ApJ...793..122K} $^{3}$ \\
    74 & Pub(Ko) & $-$ & Unbound probability in the potential model in \citet{2010ApJ...712..260K} $^{3}$ \\
    75 & Pub(P) & $-$ & Unbound probability in the potential model in \citet{1990ApJ...348..485P} $^{3}$ \\
    76 & Pub(M) & $-$ & Unbound probability in the potential model in \citet{2015ApJS..216...29B} $^{3}$ \\
    77 & Pub(X) & $-$ & Unbound probability under the potential model in \citet{2008ApJ...684.1143X} $^{3}$ \\
    78 & Origin\_Class & $-$ & Origin classification including ``HVS'', ``HRS'', ``RS'' and ``OUT'' $^{3}$ \\
    79 & T$_{\rm eff}$\_LASP & K & Effective temperature estimated by the LASP $^{3}$ \\
    80 & e\_T$_{\rm eff}$\_LASP & K & Uncertainty on Teff$\_$LASP $^{3}$ \\
    81 & log~$g$\_LASP & dex & Surface gravity estimated by the LASP $^{3}$ \\
    82 & e\_log~$g$\_LASP & dex & Uncertainty on log$(g)\_$LASP $^{3}$ \\
    83 & [Fe/H]\_LASP & dex & Metallicity estimated by the LASP $^{3}$ \\
    84 & e\_[Fe/H]\_LASP & dex & Uncertainty on [Fe/H]$\_$LASP $^{3}$ \\
    85 & T$_{\rm eff}$\_DD-Payne & K & Effective temperature estimated by the method of data-driven Payne $^{3}$ \\
    86 & e\_T$_{\rm eff}$\_DD-Payne & K & Uncertainty on Teff$\_$DD-Payne $^{3}$ \\
    87 & log~$g$\_DD-Payne & dex & Surface gravity estimated by the method of data-driven Payne $^{3}$ \\
    88 & e\_log~$g$\_DD-Payne & dex & Uncertainty on log$(g)\_$DD-Payne $^{3}$ \\
    89 & [Fe/H]\_DD-Payne & dex & Metallicity estimated by tthe method of data-driven Payne $^{3}$ \\
    90 & e\_[Fe/H]\_DD-Payne & dex & Uncertainty on [Fe/H]$\_$DD-Payne $^{3}$ \\
    91 & Alpha\_Fe\_DD-Payne & dex & $\alpha$ abudance estimated by the method of data-driven Payne $^{3}$ \\
    92 & e\_Alpha\_Fe\_DD-Payne & dex & Uncertainty on Alpha$\_$Fe$\_$DD-Payne $^{3}$ \\
    93 & conservative\_result & $-$ & HiVelSCs satisfy all more conservative criteria in Section$~$\ref{sec: other_discussion} $^{3}$ \\
\hline
\end{longtable}
\footnotesize[1]{~For unbound HiVelSCs and bound HiVelSCs with Z$_{\rm max} > 200$ kpc, we do not provide eccentricity $e$ and the maximum height above the Galactic disk $Z_{\rm max}$.}
\footnotesize[2]{~``$-$'' represents null.}

\acknowledgments

We would like to thank an anoymous referee for helpful comments and suggestions. We thank Bovy Jo, Fu Xiao-Ting, Du Bing, Yan Hong-Liang, Shi Jian-Rong, Hou Wen, Chen Jian-Jun,Yuan Hai-Long, Guo Yan-Xin, Xue Xiang-Xiang, and Zhang Lan for helpful discussions. This work is supported by National Key R$\&$D Program of China (No. 2019YFA0405502), the National Natural Science Foundation of China (Nos. U1931209, 11873056, 11991052,
11988101 and 11890694), and the National Key Program for Science and Technology Research and Development (Grant No. 2016YFA0400704). YST is supported
by the NASA Hubble Fellowship grant HST-HF2-51425.001 awarded by the Space Telescope Science Institute. Cultivation Project for LAMOST Scientific
Payoff and Research Achievement of CAMS-CAS. Guoshoujing Telescope (the Large Sky Area Multi-Object Fiber Spectroscopic Telescope, LAMOST) is a National
Major Scientific Project built by the Chinese Academy of Sciences. Funding for the project has been provided by the National Development and Reform
Commission. LAMOST is operated and managed by the National Astronomical Observatories, Chinese Academy of Sciences.

\end{document}